\documentclass[aps,prd,preprint,nofootinbib,superscriptaddress]{revtex4-2}
\usepackage{amsfonts}
\usepackage{amsmath}
\usepackage{amsthm}
\usepackage{amssymb}
\usepackage{mathtools}
\usepackage{enumitem}
\usepackage{textcomp}
\usepackage{gensymb}
\usepackage{hyperref}

\usepackage{bm}

\usepackage{xcolor}
\usepackage{dsfont}

\usepackage{lmodern}

\usepackage{graphicx}
\graphicspath{ {figures/} }
\usepackage{tipa}

\DeclareMathOperator{\Tr}{Tr}

\newcommand{\tetrad}{\text{\textschwa}}

\begin{document}
\title{
	{Canonical aspects of pregeometric vector-based first order gauge theory}
}
\author{Priidik Gallagher}
\email{priidik.gallagher@ut.ee}
\affiliation{Laboratory of Theoretical Physics, Institute of Physics, University of Tartu, W. Ostwaldi 1, 50411 Tartu, Estonia}

\begin{abstract}
	A recently proposed pregeometric auxiliary vector mediated gauge theory is studied in its canonical domain, by performing the Legendre transform on a curved background and by considering its covariant phase space, with further application to duality. The constraints become differential equations, but the Dirac-Bergmann algorithm appears consistent with electromagnetic degrees of freedom, metric background permitting. Solving the consistency conditions provides a preferred direction in an intermediary form of spontaneous symmetry breaking. In parallel, the covariant phase space defines the symplectic structure, and establishes the conserved currents and quantum phenomenology with the generated background. The formalism immediately allows to study the parent path integrals of dual theories, with quartic Proca to Kalb-Ramond inequivalence and Maxwell-Chern-Simons path integral consistency as practical applications, while the differential geometry gives a global description of the issue of Yang-Mills duality rotations.  Properties of degenerate metric geometry are discussed throughout, and the viability of inherent backgrounds leads into the fundamental question of background independence in all physical theories.
\end{abstract}

\maketitle
\newpage

\section{Introduction}
\label{sec:introduction}

Although the two common approaches of defining a physical theory, the Lagrangian and the Hamiltonian, are generally equivalent in results and capacity for a large class of common theories, either method has served separate usefulness and provided their own set of issues, so a holistic treatment studies both formulations. This is likewise the aim here, to continue the study of the author's work on pregeometric gauge theories from~\cite{Gallagher:2022kvv}, earlier provided in an action formulation, now in their Hamiltonian, canonical domain. This is performed in the standard, non-covariant method using the Legendre transform on a curved background, but also through the covariant phase space. Either approach provides some assorted results, for the theories of interest, for the general aims of theory development in pregeometry and degenerate backgrounds, as well as some general clarification for the procedures operated. For particular results, the canonical procedures prove very apt in studying constraints and degrees of freedom, and as a simple foray into quantum properties. Overall, it adds confirmation (or, rather countermand) whether a theory is well defined.

The formalism itself immediately allows to study duality. Indeed, the first order form of gauge theory, where the constitutive law has been made dynamical, precisely defines a suitable parent action and path integral used to derive dual theories. There are several important points to clarify. First, the path integral allows to verify symplectic claims, but furthermore, as will be emphasized, duality is crucially rather an identity of the master system, which can then be defined in a variety of (dual) ways. In the case of basic Proca and Kalb-Ramond theories, this identity is in opposition to the non-duality implication of~\cite{Hell:2022}. Either theory is derived from the same basic parent path integral, while it is now shown that quartically self-interacting modifications are \emph{not} dual to each other, thus limit claims are not necessarily coincident. Similarly, this identity is visible in Maxwell-Chern-Simons theory as in~\cite{Armoni:2022xhy}, as the field shift does not change the parent path integral. But as will be shown, the field shift is also consistent with direct Gaussian integration, thus not a unique method, provided vacuum consistency is properly handled. As a particular example, Proca-Chern-Simons is investigated. For this purpose, the Hamiltonian analysis formalism, and particularly the differential geometry of the $3+1$ split presented, immediately proceeds to dimensional reduction. Finally, a simple, global description to the inconsistency of Yang-Mills duality rotations~\cite{Deser:1976iy} is given, as it follows immediately from differential geometric results. A scalar multiple of curvature is not necessarily itself a curvature of some other connection. Curvatures do not form a homogeneous space. Therefore, generally there can be no transformations to a ``magnetic'' Yang-Mills theory.

Initially, the primary interest is in the vector $\phi^a$ theory of~\cite{Gallagher:2022kvv} due to its nontrivial phenomenology. This is the so-called isokhronon theory, which was developed for the express purpose of analogy and compatibility with the \emph{khronon} gravitational theory of~\cite{Zlosnik:2018}. To some length this similarity is rather extensive, but with some important torsion- and background-related problems, as will be discussed. Nevertheless, it has proven to be a surprisingly interesting model outside of simple mathematical peculiarity and physical aspiration, far more than such a problematic theory should have. There's been some recent interest in various first order models, having been brought up in gravity~\cite{Rasanen2023Palatini} and (topological) gauge theory ~\cite{husain2023general}, but more curiously there was also a recent suggestion of a fundamental background ambiguity in quantum field theory, per applicability of the Gauss law constraint~\cite{kaplan:2023a, kaplan:2023b}. The paper~\cite{casadio2024relaxation} provides a great detail of further background into these ideas as well. The $\phi^a$ modification to usual electromagnetism gives one classical analogue of the phenomenon, in addition to similar behaviour in the theory of khronon gravity~\cite{Zlosnik:2018}. It will not be overly difficult to propose other (both simpler and better behaving) similar models with background ambiguity\footnote{Also for further comparison, higher derivative (gauge) theories can similarly provide additional degrees of freedom (and mostly are not physically viable), see e.g.~\cite{Crisostomi:2017aim}.}. Overall this is the more general question of background independence in physics, outside of the meaning it has in gravity. For comparison, taking this a step further would be questioning why not to add any arbitrary background current as long as the transformation properties are satisfied per Noether's theorems and gauge invariance.

For the auxiliary vector gauge theory in question, this ambiguity manifests as the constraints become differential equations --- a phenomenon similar to what was discussed in~\cite{dambrosio:2023} as an issue that could break the Dirac-Bergmann algorithm. So, the theory under study is an explicit (counter)example for the Dirac-Bergmann algorithm in field theory with differential equation constraints, although here, at least superficially, the algorithm still appears to provide the expected number of \emph{local} degrees of freedom for electromagnetism. The vacuum excitation is then situated in global degrees. For the model in question, this works only when the background is compatible, although the general suggestion is further-reaching.

Studying the constraint and their consistency conditions also provides what appears to be spontaneous Lorentz symmetry breaking, as solving the system requires introducing an explicit and arbitrary direction, thus introducing a preferred direction and apparently breaking Lorentz symmetry. Note this must classically be in an intermediary form, however, because if the initial data also properly transform under symmetry, there is no explicit violation of the symmetry, except in an ad hoc discerning of observer and particle-like transformations. Nevertheless, this is now visible in multiple ways. In the Lagrangian approach this proceeds through integrating the equations of motion, which introduces an integration constant vacuum excitation background. The shift symmetry of the theory, as discussed in~\cite{Gallagher:2023}, \emph{also} reminds of the shift symmetry of the Nambu-Goldstone realization, although requires closer inspection. Spontaneous Lorentz symmetry breaking was a prominent feature of the khronon gravity theory~\cite{Zlosnik:2018}.

These conclusions are reached primarily through a non-covariant Hamiltonian analysis of the theory. The Legendre transformation and constraint algebra analysis have been thoroughly studied throughout literature, including on curved spacetime. Here also, as far as is reasonably possible and meaningful, the constraint analysis is performed on a curved background, using the differential form method as in~\cite{Okolow_2012}. Generally establishing the Hamiltonian formalism requires a globally hyperbolic manifold: some comment on possible deviations or extensions is provided throughout.

The non-covariant analysis benefits in its methods being explicit, even if not necessarily easy. However, there exist covariant alternatives to define the phase space of a physical theory. These covariant phase space \emph{methods}, as there are in principle multiple possibilities, have recently had a surge in interest~\cite{Harlow:2019yfa, Freidel:2020xyx, Margalef:2021, ciambelli:2023, gieres2023covariant, assanioussi:2023, delValleVaroGarcia:2022ceq}. In this paper, we also look at what is known as \emph{the} covariant phase space. This name is justified as the formalism provides a symplectic structure on the phase space, which is now established by the solutions of the equations of motion, thus retaining (implicit) covariance. The variation can be thought of as the central procedure for establishing the phase space, thus also immediately suggesting to study the conserved currents. In part, they were already reported in~\cite{Gallagher:2023}, but some expansion of the results and some of the details of Noether's theorems are collected here, in particular related to the applicability for local or gauge symmetries and in the presence of backgrounds. Similarly, the symplectic structure, being immediately related to the Poisson brackets, serves as the basis for various canonical quantization procedures. This provides a very simple probe of quantum properties: symplectic equivalence would suggest quantization equivalence, if the algebro-geometric arena can be taken in isolation, while the solution space also gives a \emph{quantum} characterization of the isokhronon theory --- which appears to be quantum gauge theory on a \emph{classical} electromagnetic background. In particular, handling the integration constant background wasn't entirely clear on the quantum level earlier. In principle, the quantum theory does not suggest to be substantially different from any particle theory in strong backgrounds, which are already well understood~\cite{Fedotov:2023}.

The term \emph{pregeometric} is used to refer to theories which allow a consistent zero metric phase, like a vanishing metric ground state. In~\cite{Zlosnik:2018} it is a vector (the so-called \emph{khronon}) which generates the coframe, thus the metric. A related understanding is that the metric and related concepts are an emergent phenomenon, generated by more fundamental fields. As formulated, these are distinct concepts, although they often touch on the same problems. So let us offer another, perhaps more practical formulation or aspiration: it is the idea to require all physical theories to support a topological phase and a consistent transition. The references~\cite{Gallagher:2022kvv, Gallagher:2023} provide more discussion and examples in the literature, but the lay of the theoryscape and ideas is rather varied, from quantum effects or Planck-scale geometry to black hole singularities, and the main obstacle is often gauge theory (rather than spinor matter), intricately dependent on the metric. As it currently stands, the field is still open, and there is at least some hope of developing the standard formulations of physics further when adopting this different perspective, but there's still much work to be done. For terminological completion, the \emph{premetric} programme refers to an axiomatic derivation of (gauge) theory structure: see~\cite{Hehl_Obukhov:2003} for a thorough discussion of electromagnetism in this view.

The article is structured into three major sections. Section~\ref{sec:noncovariant} begins by introducing the standard Hamiltonian formalism through the Legendre transformation on a curved background, and then performing the constraint analysis of the pregeometric gauge theory in question, with comments of wider applicability throughout. Section~\ref{sec:covariant} studies the covariant phase space in application to the theory of interest. The basic operation, the variation of the action, is very simple but the implications of it are somewhat more wide-reaching, thus the covariant phase space, the conserved currents, and some relevant quantization procedures are explained in a little more detail beforehand. Finally, section~\ref{sec:duality} continues application to duality. The path integral of the isokhronon theory is studied, alongside the duality of Proca-Kalb-Ramond and Maxwell-Chern-Simons. Finally, the differential geometry of duality rotations is studied. Implicitly, the setting is on a 4-dimensional smooth manifold, unless stated otherwise, with the metric signature $(-,+,+,+)$ and the Levi-Civita symbol $\epsilon_{0123}=+1$, although this does not actually significantly come up in the formalism. Fundamental constants are normalized, in particular $c=\hbar=1$.

\section{Noncovariant Hamiltonian analysis}
\label{sec:noncovariant}

\subsection{Canonical noncovariant geometry}
\label{subsec:noncovariant_geometry}

Any book including canonical quantization or the ADM formalism will provide an introduction to the Hamiltonian formulation of field theory. For the purposes here, the analysis is throughout written in differential forms, as presented in e.g.~\cite{Okolow_2012}. This is simply a change in description, as differential forms compose the antisymmetric subspace of tensors, while vector- and tensor-valued forms provide the remainder. The benefit of differential forms is the consistency of description and simplicity of geometric identities and structures, particularly attractive for theory construction, but not necessarily in direct calculations.

The standard Legendre transformation, in gravity~\cite{Arnowitt:1962hi, bojowald:2010} and in field theory~\cite{Fulling:1989nb, Hehl_Obukhov:2003, Rothe_Rothe:2010}, assumes a globally hyperbolic manifold, thus the manifold topology is $M=\mathbb{R}\times\Sigma$, a sequence of spatial hypersurfaces $\Sigma$, parametrized by a monotonously increasing time variable $\sigma$~\cite{Bernal:2003jb}. The direction of time vector field $n$, a congruence of observer worldlines, is defined by the interior product, resp. the Lie derivative via Cartan's homotopy formula,
\begin{equation}
	n\lrcorner\mathrm{d}\sigma=\mathcal{L}_n\sigma=1,
\end{equation}
so that a $p$-form $\alpha$ can be split with respect to $n$ into longitudinal (temporal) $\alpha_\perp$ and transverse (spatial) $\underline{\alpha}$ components, respectively
\begin{align}
	{}^\perp\alpha&=\mathrm{d}\sigma\wedge\alpha_\perp,\quad\alpha_\perp=n\lrcorner\alpha,\\
	\underline{\alpha}&=(\mathds{1}-{}^\perp)\alpha=n\lrcorner(\mathrm{d}\sigma\wedge\alpha),\quad n\lrcorner\underline{\alpha}=0.
\end{align}
In other words
\begin{equation}
	\alpha=\mathrm{d}\sigma\wedge\alpha_\perp+\underline{\alpha}.
\end{equation}
The spatial part of the form is situated on the spatial hypersurfaces $\Sigma$, and properly defined via the pullback through the embedding $\varphi:\Sigma\to M$, see e.g.~\cite{Hohmann:2020}. For simplicity, this will be left implicit in the definition, using the projection by $n$. There will be no ambiguity as the pullback behaves well under exterior calculus.

This split defines an adapted local coordinate system $x^a=(\sigma,x^I)$, with
\begin{equation}
	e^a=(e^0,e^I)=(\mathrm{d}\sigma,\underline{\mathrm{d}x^a})
\end{equation}
as the basis of the cotangent space. The corresponding natural frame basis is
\begin{equation}
	\tetrad_a=(n,\tetrad_I),
\end{equation}
with $e^a(\tetrad_b)=\delta^a_b$. In particular,
\begin{align}
	e^a_\perp&=\delta^a_0,\\
	\underline{e^a}&=\delta^a_I e^I,
\end{align}
which is convenient for explicit calculations.

The exterior derivative decomposes as
\begin{equation}\label{eq:ext_der_split}
	{}^\perp(\mathrm{d}\alpha)=\mathrm{d}\sigma\wedge(\mathcal{L}_n\underline\alpha - \underline{\mathrm{d}}\alpha_\perp),\quad\underline{\mathrm{d}\alpha}=\underline{\mathrm{d}}\,\underline{\alpha},
\end{equation}
as the Lie derivative along a vector field $n$ can be expressed using Cartan's homotopy formula
\begin{equation}
	\mathcal{L}_n\alpha=n\lrcorner\mathrm{d}\alpha + \mathrm{d}(n\lrcorner\alpha).
\end{equation}
The Lie derivative of the transversal part $\underline{\alpha}$ with respect to the vector $n$ is abbreviated by a dot,
\begin{equation}
	\underline{\dot{\alpha}}=\mathcal{L}_n\underline{\alpha},
\end{equation}
since it is the time derivative of the corresponding quantity. Of note is that when the Lagrangian is restricted to natural operations on forms, the Lie derivative of $\alpha_\perp$ does not appear, so from the viewpoint of canonical formalism, $\alpha_\perp$ can be immediately seen as a Lagrange multiplier. Then $\underline{\alpha}$ is the only dynamical quantity.

There are several notions of change in time in a curved spacetime~\cite{Thorne_MacDonald:1982a, Thorne_MacDonald:1982b}, especially apparent for vector-valued forms or any field in nontrivial representations of their structure group. The principal concept for Hamiltonian analysis still remains the non-covariant Lie-dragging of the object on the base manifold, along the direction of time vector field. This is the method used for Hamiltonian analysis in theories of gravity, similarly in lattice calculations, see e.g.~\cite{Lewis2020classical}, and it remains true in classical and quantum field theory on curved spacetime.

Parallel transport in fiber bundles, vector bundles in specific, necessitates the additional concept of connection, for a connection form $\omega$ and its exterior covariant derivative $\mathrm{D}$. This also provides a ``covariant Lie derivative''
\begin{equation}
	\mathfrak{L}_n\alpha=\mathrm{D}n\lrcorner\alpha + n\lrcorner\mathrm{D}\alpha,
\end{equation}
which comes up occasionally, see e.g.~\cite{Gronwald:1998ag,Obukhov_Rubilar:2006}. Another concept is the longitudinal projection $(\mathrm{D}\alpha)_\perp = n\lrcorner\mathrm{D}\alpha$. The relations are clear in 3+1 form, where the covariant Lie derivative $\alpha$ decomposes as
\begin{equation}
	\begin{aligned}
		\mathfrak{L}_n\alpha&=\mathrm{d}\sigma\wedge(\mathcal{L}_n\alpha_\perp + \omega_\perp\alpha_\perp) + \mathcal{L}_n\underline{\alpha} + \omega_\perp\underline{\alpha}=\\
		&=\mathrm{d}\sigma\wedge\mathfrak{L}_n\alpha_\perp + \mathfrak{L}_n\underline{\alpha},
	\end{aligned}
\end{equation}
while the exterior covariant derivative splits as
\begin{equation}
	\begin{aligned}
		\mathrm{D}\alpha&=\mathrm{d}\sigma\wedge(\mathcal{L}_n\underline{\alpha} + \omega_\perp\underline{\alpha} - \underline{\mathrm{d}}\alpha_\perp - \underline{\omega}\wedge\alpha_\perp) + \underline{\mathrm{d}}\,\underline{\alpha} + \underline{\omega}\wedge\underline{\alpha}\equiv\\
		&\equiv\mathrm{d}\sigma\wedge(\mathfrak{L}_n \underline{\alpha} - \underline{\mathrm{D}}\,\alpha_\perp) + \underline{\mathrm{D}}\,\underline{\alpha}.
	\end{aligned}
\end{equation}
Furthermore, introducing derivatives with respect to proper time would measure change with respect to the observer's internal clock. So altogether there appear many different concepts of change in time, in particular: $\mathcal{L}_n$, $\mathfrak{L}_n$ and $n\lrcorner\mathrm{D}$. The difference is simply in the geometric meaning: the Lie derivative drags the object along the base manifold, while the covariant derivatives correct the change to remain horizontal, thus gauge covariant. The projection via the interior product measures the total covariant change in the given direction. But to emphasize again, the basic Lie derivative remains the basic notion of derivative in canonical analysis. This is naturally visible, when defining the Lie derivative through the change of a tensor along the flow of a vector field, using the pullback.

Altogether, this formalism applied to the action provides
\begin{equation}
	S[\alpha]=\int L[\alpha,\mathrm{d}\alpha]=\int\mathrm{d}\sigma\wedge L_\perp,
\end{equation}
where the canonical momenta is defined by variation
\begin{equation}
	\delta_{\underline{\dot{\alpha}}}{L}_\perp=\delta \underline{\dot{\alpha}}\wedge \pi_\alpha,
\end{equation}
so the canonical Hamiltonian is defined by the Legendre transformation
\begin{equation}
	H_c=\int\underline{\dot{\alpha}}\wedge\pi_{\alpha}-L_\perp,
\end{equation}
with an implicit sum over all fields $\alpha$. The variational bicomplex and jet bundles would provide a more sophisticated treatment of the geometric theory Lagrangian mechanics, see e.g.~\cite{Anderson:1992}. But for the purposes here, functional constraints as spatial integrals with Lagrange multipliers are to be supplemented. The constraints enforcing the definitions of the momenta are to be immediately supplemented. Altogether, this provides the primary Hamiltonian. The field-theoretical Poisson brackets for functionals $F$ and $G$ read as
\begin{equation}
	\{F,G\}=\int\bigg(\frac{\delta F}{\delta \underline{\alpha}}\wedge\frac{\delta G}{\delta \pi_{\alpha}} - \frac{\delta G}{\delta \underline{\alpha}}\wedge\frac{\delta F}{\delta \pi_{\alpha}}\bigg),
\end{equation}
and allow studying the constraint algebra. In particular, this allows operating the Dirac-Bergmann algorithm, and classifying constraints as primary, secondary, or first-class and second-class. Note that the direct continuation to vector- and tensor-valued forms includes contractions through the metric. Overall, this is simply a translation of the Hamiltonian analysis procedure to the language of differential forms.

Now, in a theory with a possibly degenerate metric $g$ background, the manifold $M$ can be separated into metric regions $M_g$, with nonzero $g$, and topological regions $M_\text{Top}$, where $g$ vanishes. In practice the details of degeneracy in the physical model are often very theory-dependent; for comparison, quantum properties and the zero ground state of the metric~\cite{Witten:1988npb, Horowitz:1991, Giddings:1991, Banados_2007} versus the singularity of black holes or gravitational instantons~\cite{DAuria:1981ddz, Bengtsson:1993}. But when restricting to at least effective backgrounds, the standard mathematical and physical apparatus,  including the canonical structure, remains locally valid in the regions equipped with a standard pseudo-Riemannian metric. The effect of the degenerate regions is in introducing topological contention to the manifold. This is not entirely unlike bimetric theory, such as the classic proposal by Rosen~\cite{Rosen:1973,Rosen:1975}, where there are several notions (and interpretations) of the metrics in contention, yet provides interesting phenomenology~\cite{Schmidt-May:2016}.

Geometry-wise, the topological region $M_\text{Top}$ loses the light cone causal structure generated by the metric. Only topological notions remain, and the geometry isn't notably different from that of a bare manifold. An observer cannot discern the passage of time or extent of space. In the manifold and physical field theory sense, the topological regions viewed from outside, that is from the metric regions, manifest as either discontinuities, topological defects (or features), singularities, or just the ``interior'' of a single spacetime point. As such, naming nonmetric regions, i.e.\ degenerate regions of $g=0$, as topological is justified --- the nontrivial topological effects of such regions are of primary interest, compared to standard metric topology. Classically, zero distant points can often be factored out to a single equivalence class, a single topological feature, cf. the identified metric construction in pseudometric spaces\footnote{Here in the sense of the mathematical generalization of metric spaces, which permit zero distance points, i.e.\ points do not have to be metrically distinguishable. So, not to be conflated with the pseudo-Riemannian metric.}. Nevertheless, (smooth) limits of the metric tensor can still provide nontrivial asymptotics. Overall a zero metric, thus also a zero geodesic distance implies points which are not just close, but metrically coincide. It is inevitably singular geometry, even if there is some desire in allowing for it.

In dynamical terms, a topological region can nontrivially only facilitate a topological (quantum) field theory, as any Lagrangian term which includes the metric would simply vanish. Any term involving the inverse metric would become singular. In other words, and without invoking the Lagrangian, the lack of light cone structure simply does not allow separating any sense of change in time from change in space. So, in these topological regions, the pregeometric theories desire to propose a phase transition, compared to singularities in standard gauge theory. Overall the physical effects remain theory-dependent, see e.g.~\cite{Tucker:1995,Gratus:1996mr}.

In canonical terms, topological quantum field theories have a zero (or undefined) Hamiltonian in general, since a bare manifold has no consistent sense of time~\cite{Witten:1988ze, Witten:1988hf, Atiyah:1989vu, Birmingham_Blau_Thompson:1990}. Although they do not retain local degrees of freedom (by definition) it is still possible to have a nontrivial effective theory on the boundary, see~\cite{Corichi_Vukasinac:2019} and the references therein. Generally and in practical terms, the main physical interest is in finding any backreaction into the metric regions of spacetime, which host our observed reality.

Focusing on the canonical structure, vector fields as a concept are independent of the metric, so, topology permitting, it is not strictly impossible to smoothly extend the direction of time vector field inside the topological region. However, this would enforce a preferred direction of time, and extend the existence of a Lorentz metric inside the topological region, see e.g.~\cite{oneill:1983} for details, and also~\cite{Wald:1984}. This metric would not necessarily coincide with the one defining the topological region, but the primary issue is that there is no and should be no preferred concept of an observer worldline through such regions: the entire set $\text{Vect}(M_\text{Top})$ of vector fields in the topological region are equally permitted, and shouldn't be considered separately. On a more pragmatic level, such regions run into trouble with the requirement of global hyperbolicity in the standard understanding, which is largely also a topological requirement. This does not necessarily mean that differential equations cannot be solved, but rather that the degenerate behaviour has to be looked into by hand.

Overall, standard and topological quantum field theories are separately well understood. The following focuses only on the local canonical structure in the metric regions, where such a consideration is sensible and has not been yet verified for the theory of interest. Behaviour around degeneracies should be verified separately, even if there is hope that the dynamical behaviour is (somewhat) improved. This could be done in perturbations around the degeneracy, but also, as proposed in the conclusion, metrically contracting a topological feature. Either promise an interesting study.

\subsection{Constraint analysis}
\label{subsec:constraint}

Let us study the electromagnetic isokhronon $\phi^a$, a Lorentz vector valued in the adjoint of the gauge group, presented in~\cite{Gallagher:2022kvv} to be
\begin{equation}\label{eq:isokhronon_action_basic}
	S_{\phi}=\frac{1}{2\mu_g}\int\bigg(\frac{1}{2}\epsilon_{abcd} \mathrm{D}\phi^a\wedge e^b\wedge \mathrm{D}\phi^c\wedge e^d + \eta_{ab} \mathrm{D}\phi^a\wedge e^b\wedge F\bigg).
\end{equation}
Let $\frac{1}{2\mu_g}=1$, as the coupling constant is currently irrelevant. The pseudo-coframe $u^a$ (being the analogue of $\mathrm{D}\phi^a$) and the linear transformation $G_{ab}$ (cf. $\mathrm{D}_{[a}\phi_{b]}$) variants are comparatively trivial, and not substantially different from the analysis of standard 1st order electromagnetism~\cite{Sundermeyer:1982, Kiriushcheva_Kuzmin_McKeon:2012}. Yang-Mills theory would use non-Abelian gauge groups, but the calculations are similar in both Lagrangian and Hamiltonian terms, when the trace is consistently handled.

Lagrangian dynamics for the electromagnetic theory is described by the equations of motion
\begin{subequations}
\label{eq:phi_LagrangianEoM}
\begin{align}
	\mathrm{D}(\mathrm{D}\phi_a\wedge e^a) &= 0,\label{eq:phimaxwell_basic}\\
	\mathrm{D}(\epsilon_{abcd}\mathrm{D}\phi^b\wedge e^c\wedge e^d + e_a\wedge F)&=0,\label{eq:auxiliar_constitutive_basic}
\end{align}
\end{subequations}
the first being the prototype inhomogeneous Maxwell (resp. Yang-Mills) equation and the second being the constitutive law of premetric fare, essentially a covariant field excitation constraint. The latter can be integrated to
\begin{equation}\label{eq:auxiliar_constitutive_integrated}
	\epsilon_{abcd}\mathrm{D}\phi^b\wedge e^c\wedge e^d + e_a\wedge F=X_a,\ \mathrm{D}X_a=0,
\end{equation}
which implies
\begin{equation}
	\mathrm{D}\phi_a\wedge e^a = *F + \frac{1}{2}(*X_a)\wedge e^a.
\end{equation}
Therefore $\phi^a$ can be integrated out, classically yielding electromagnetism and a vacuum excitation. Similarly, when a specific $X_a$ is chosen,~\eqref{eq:auxiliar_constitutive_integrated} is no longer shift symmetric, compared to~\eqref{eq:auxiliar_constitutive_basic}. Note, however, that this integration \emph{only} makes sense if torsion $T^a=\mathrm{D}e^a$ is nonvanishing, as otherwise the second term of~\eqref{eq:auxiliar_constitutive_basic} loses meaning.

The 3+1 split of these equations will prove to be useful later on. The equations of motion~\eqref{eq:phi_LagrangianEoM} are respectively
\begin{subequations}
\begin{gather}
	\mathrm{d}\sigma\wedge(\mathcal{L}_n(\underline{\mathrm{D}}\phi_a\wedge\underline{e^a})
	-\underline{\mathrm{D}}(\mathfrak{L}_n\phi_a \underline{e^a} - \underline{\mathrm{D}}\phi_a e_\perp^a))
	+ \underline{\mathrm{D}}(\underline{\mathrm{D}}\phi_a\wedge\underline{e^a})=0,\label{eq:31_proto_inhom}\\
	\begin{multlined}
	\mathrm{d}\sigma\wedge(\mathfrak{L}_n(\underline{\epsilon_{abcd}\mathrm{D}\phi^b\wedge e^c\wedge e^d + e_a\wedge F}) \\
	{}-\underline{\mathrm{D}}(\epsilon_{abcd}((\dot{\phi}^b+\omega_\perp{}^b{}_i\phi^i)\underline{e^c}\wedge\underline{e^d}+2e_\perp^b\underline{\mathrm{D}}\phi^c\wedge\underline{e^d} )
	+ \eta_{ab}e_\perp^b\underline{F} - \eta_{ab}\underline{e^b}\wedge(\underline{\dot{A}} - \underline{\mathrm{d}} A_\perp))=0.\label{eq:31_unintegrated_constitutive}
\end{multlined}
\end{gather}
\end{subequations}
After integrating the total covariant derivative, the auxiliary equation becomes
\begin{multline}\label{eq:31_constitutive_law}
	\mathrm{d}\sigma\wedge(\epsilon_{abcd}((\dot{\phi}^b+\omega_\perp{}^b{}_i\phi^i)\underline{e^c}\wedge\underline{e^d}+2e_\perp^b\underline{\mathrm{D}}\phi^c\wedge\underline{e^d} )
	+ \eta_{ab}e_\perp^b\underline{F} - \eta_{ab}\underline{e^b}\wedge(\underline{\dot{A}} - \underline{\mathrm{d}} A_\perp) -   X_{\perp a})\\
	{}+\underline{\epsilon_{abcd}\mathrm{D}\phi^b\wedge e^c\wedge e^d + e_a\wedge F} - \underline{X_a}=0.
	\end{multline}
The 3-form $X_a$ being covariantly constant means
\begin{equation}
	\mathrm{D}X_a=\mathrm{d}\sigma\wedge(\underline{\dot{X}_a} + \omega_{\perp a}{}^b\underline{X_b} - \underline{\mathrm{D}}X_{\perp a}) = 0.\label{eq:covariant_constant}
\end{equation}
Importantly, the spatial part of~\eqref{eq:31_constitutive_law} provides the meaning of the integration constant $\underline{X_a}$ as
\begin{equation}
	\underline{X_a}=\underline{\epsilon_{abcd}\mathrm{D}\phi^b\wedge e^c\wedge e^d + e_a\wedge F}.
\end{equation}
which can be matched with the isokhronon canonical momentum later on. The longitudinal component $X_{\perp a}$ does not have a definition as neat as the former, although the longitudinal component implies
\begin{equation}
	X_{\perp a}=\epsilon_{abcd}((\dot{\phi}^b+\omega_\perp{}^b{}_i\phi^i)\underline{e^c}\wedge\underline{e^d}+2e_\perp^b\underline{\mathrm{D}}\phi^c\wedge\underline{e^d} )
	+ \eta_{ab}e_\perp^b\underline{F} - \eta_{ab}\underline{e^b}\wedge(\underline{\dot{A}} - \underline{\mathrm{d}} A_\perp),
\end{equation}
and it's likewise constrained by~\eqref{eq:covariant_constant}.

Returning to Hamiltonian analysis, the canonical Hamiltonian reads
\begin{equation}
	H_c=\int
	\begin{aligned}[t]
		\Big[{}-{}&\epsilon_{abcd}\omega_\perp{}^a{}_i\phi^i\underline{e^b}\wedge\underline{\mathrm{D}}\phi^c\wedge\underline{e^d}-\epsilon_{abcd}e^a_\perp\underline{\mathrm{D}}\phi^b\wedge\underline{\mathrm{D}}\phi^c\wedge\underline{e^d}\\
		{}-{}&\eta_{ab}\omega_\perp{}^a{}_i\phi^i\underline{e^b}\wedge\underline{\mathrm{d}}\,\underline{A}+\eta_{ab}\underline{\mathrm{D}}\phi^a e^b_\perp\wedge\underline{\mathrm{d}}\,\underline{A}\Big],
	\end{aligned}
\end{equation}
while the primary Hamiltonian appends the constraint functionals
\begin{subequations}
	\begin{align}
		C_{\pi_\phi}&=\int\eta_{ij}\lambda^i_{\pi_\phi}(\pi^j_{\phi}-\epsilon^j{}_{bcd}\underline{e^b}\wedge\underline{\mathrm{D}}\phi^c\wedge\underline{e^d}-\underline{e^j}\wedge\underline{\mathrm{d}}\,\underline{A}),\\
		C_{\pi_A}&=\int\lambda_{\pi_A}\wedge(\pi_{A}-\eta_{ab}\underline{\mathrm{D}}\phi^a\wedge\underline{e^b}),\\
		C_{A_\perp}&=-\int\eta_{ab}\underline{\mathrm{D}}\phi^a\wedge\underline{e^b}\wedge\underline{\mathrm{d}}A_\perp
\sim
\int A_\perp\underline{\mathrm{d}}(\eta_{ab}\underline{\mathrm{D}}\phi^a\wedge\underline{e^b}),
	\end{align}
\end{subequations}
to a total of
\begin{equation}
	H_p=H_c+C_{\pi_\phi} + C_{\pi_a} + C_{A_\perp}.
\end{equation}
With the prescient knowledge from having already done the calculation, a secondary constraint is picked up from maintaining the definition of the momentum $\pi^a_\phi$ in time,
\begin{equation}
	\{C_{\pi_\phi},H_p\}\equiv0	\Rightarrow\begin{multlined}[t]
		\underline{T_u}\wedge\lambda_{\pi_A}-\omega_\perp{}^a{}_u(\epsilon_{abcd}\underline{\mathrm{D}}\phi^b\wedge\underline{e^c}\wedge\underline{e^d}+\eta_{ab}\underline{e^b}\wedge\underline{F})\\
		{}+\underline{\mathrm{D}}(\epsilon_{aucd}\omega_\perp{}^a{}_i\phi^i\underline{e^c}\wedge\underline{e^d}
		+2\epsilon_{aucd}e^a_\perp\underline{\mathrm{D}}\phi^c\wedge\underline{e^d}
		-\eta_{ub} e^b_\perp\underline{F})=0.\label{eq:secondary_initial}
		\end{multlined}
\end{equation}
Let us postpone looking into the other constraints before handling this system of equations. In essence,~\eqref{eq:secondary_initial} is a system of four differential equations for the three components of $\lambda_{\pi_A}$. It can be projected onto four linearly independent directions, i.e.\ basis vectors; simply choosing components $u=0,1,2,3$ would be implicitly projecting on the directions $\delta^u_{0,1,2,3}$ of the orthonormal basis $\tetrad_a$. There is no invariant method of separating three equations from the fourth, extraneous one --- the split is not unique. On the linear level, any Gauss-elementary combination of the initial equations still provides an equivalent system. In other words, choosing a secondary constraint exhibits spontaneous symmetry breaking of Lorentz covariance, as it prefers a unique direction. This mirrors the initial \emph{khronon} theory of gravity~\cite{Zlosnik:2018}, for an electromagnetic (or Yang-Mills) analogy.

Let it be emphasized that here, classically, this is an \emph{intermediary} form of spontaneous symmetry breaking. The choice of $v_0$ is persistent, as is the ultimately resulting background $X_a$. A preferred background frame, as picked out to retain $v_0$ or $X_a$ invariant, would be in contradiction to relativity. Nevertheless, for Lorentz symmetry to be actually explicitly violated, $v_0$ or $X_a$ should not transform under the symmetry. This is directly the case for fixed backgrounds, see~\cite{Bluhm:2014oua, Bluhm:2023kph, Reyes:2024hqi}. It is also prudent to distinguish between observer and particle transformations, where observer diffeomorphisms necessarily act passively and covariantly on background and dynamical fields alike, while particle diffeomorphisms only affect dynamical objects\footnote{Also note Elitzur's theorem, by which only locally gauge-invariant operators can have non-vanishing expectation values. A precise description of Lorentz symmetry breaking requires greater care.}. Overall, though, the initial data can also be allowed to transform under symmetry. Thus, the choice of $v_0$ or $X_a$, on the level of classical theory, is symmetry breaking only in the weaker sense of providing a preferred reference, should it be strictly enforced. While background fields are commonly referred to as symmetry breaking, the background field would actually have to have inconsistent transformation properties or preferred frame effects for proper violation to be present.

Let $v_0^u$ define the direction of the secondary constraint, while the remaining three linearly independent vectors define equations to be solved for $\lambda_{\pi_A}$, and complete a basis of spacetime tangent vectors. Altogether let this set be $\{v_0^u,v^u_{A=1,2,3}\}$. The indices $A,B,C,...$ refer to this basis, in contrast to $I,J,K,...$ of the spatial hypersurface. The most obvious option would be a secondary constraint in the direction of time, $v_0^u=e^u_\perp$, with the orthogonal directions being spatial coordinates, but this isn't necessarily the only option. Another choice would be a direction orthogonal to spatial torsion, such that $v_0^u\underline{T_u}=0$. In the context of khronon gravity~\cite{Zlosnik:2018}, there is a canonical choice of symmetry breaking in direction of the khronon vector field, $v_0^a=\tau^a$. In this theory the identity $\tau_a T^a=\tau_a \mathrm{D}\mathrm{D}\tau^a=\tau_a R^a{}_b\tau^b$ remains gauge-invariant and well-defined, and the analysis mirrors that of the case $v_0^u\underline{T_u}=0$.

The secondary constraint is
\begin{equation}
		v_0^u\underline{T_u}\wedge\lambda_{\pi_A}
		=-v_0^u\omega_\perp{}_{ua}\pi^a_{\phi}
		+v_0^u\underline{\mathrm{D}}(\epsilon_{uabc}(\omega_\perp{}^a{}_i\phi^i\underline{e^b}\wedge\underline{e^c}
		+2e^a_\perp\underline{\mathrm{D}}\phi^b\wedge\underline{e^c})
		+\eta_{ua} e^a_\perp\underline{F})\equiv v_0^u\underline{W_u},\label{eq:secondary_basic}
\end{equation}
while $\lambda_{\pi_A}$ is to be solved from
\begin{equation}
		v^u_A\underline{T_u}\wedge\lambda_{\pi_A}
		=v_A^u\underline{W_u}\equiv\underline{W_A}.
\end{equation}
Once a direction $v_0^u$ has been chosen, there is no ambiguity in the solution of the equations. As $\{v_0^u,v^u_{A=1,2,3}\}$ is assumed linearly independent, that is a basis of 4-vectors, any \emph{other} vector $s^u$ can be split as $s^u=s^0 v_0^u + s^A v^u_A$, so no additional constraint is picked up. Solving the system for $\lambda_{\pi_A}$ implicitly assumes that $v^u_A\underline{T_u}$ does \emph{not} vanish. In particular, $v^u_A$ can not be orthogonal to torsion, and torsion $\underline{T_u}$ itself cannot vanish. Such restrictions do not apply for $v_0^u$, as it would simply change the character of the constraint~\eqref{eq:secondary_basic} to 
\begin{equation}
	v_0^u\underline{W_u}=0.\label{eq:secondary_vanishingT}
\end{equation}

Define the conjugate density
\begin{equation}
	\frac{1}{2!}\underline{T_u}{}_{IJ}\epsilon^{IJK}\equiv \underline{\tilde{T}_u}{}^K,
\end{equation}
that is
\begin{equation}
		v^u_A\underline{\tilde{T}_u}{}^K\lambda_{\pi_A}{}_K=
		\frac{1}{3!}\underline{W_A}{}_{IJK}\epsilon^{IJK}.
\end{equation}
Based on the preceding discussion, it has to be \emph{assumed} that $v^u_A\underline{\tilde{T}_u}{}^K$ is left-invertible,
\begin{equation}
	M_I{}^A(v^u_A\underline{\tilde{T}_u}{}^K)=\delta_I^K.
\end{equation}
Left-inverses exist generally for any $m\times n$ matrix, $n\leqslant m$, if its rank is $n$, and right inverses proceed similarly; in the language used here this is the assumption that $v^u_A$ are linearly independent. Therefore, the solution is
\begin{equation}
	\lambda_{\pi_A}=M_K{}^B\underline{e^K}\,\underline{\star}\,\underline{W_B}.
\end{equation}

This constraint is actually not surprising, as an analogue hides in the Lagrangian equations of motion~\eqref{eq:31_unintegrated_constitutive}. The longitudinal component of the equation loses time evolution in directions $v_0^u$ orthogonal to the vector composed of time derivative terms, such that
\begin{equation}
	v_0^a(\underline{\dot{X}_a} - \underline{\mathrm{D}}(\epsilon_{abcd}\dot{\phi}^b\underline{e^c}\wedge\underline{e^d}
	- \eta_{ab}\underline{e^b}\wedge(\underline{\dot{A}}-\underline{\mathrm{d}} A_\perp)))=0.
\end{equation}
The remainder is the constraint
\begin{equation}
	v_0^a\omega_{\perp a}{}^i\underline{X_i}-
	v_0^a\underline{\mathrm{D}}(\epsilon_{abcd}(\omega_\perp{}^b{}_i\phi^i\underline{e^c}\wedge\underline{e^d}+2e_\perp^b\underline{\mathrm{D}}\phi^c\wedge\underline{e^d} )
	+ \eta_{ab}e_\perp^b\underline{F} )=0,
\end{equation}
which is precisely the secondary constraint~\eqref{eq:secondary_vanishingT} in directions of vanishing torsion.

So altogether the secondary constraint functional
\begin{equation}
	C_2=\int \lambda_2 v_0^i(\underline{\tilde{T}_i}{}^K M_K{}^B v_B^u-\delta_i^u)\underline{W_u}\equiv\int \lambda_2 V_0^u\underline{W_u}
\end{equation}
has to be appended, with
\begin{subequations}
	\begin{align}
		V_0^u&=v_0^i(\underline{\tilde{T}_i}{}^K M_K{}^B v_B^u-\delta_i^u),\\
		\underline{W_u}&=-\omega_\perp{}_{ua}\pi^a_{\phi}
		+\underline{\mathrm{D}}(\epsilon_{uabc}(\omega_\perp{}^a{}_i\phi^i\underline{e^b}\wedge\underline{e^c}
		+2e^a_\perp\underline{\mathrm{D}}\phi^b\wedge\underline{e^c})
		+\eta_{ua} e^a_\perp\underline{F}).
	\end{align}
\end{subequations}
When projecting in directions with vanishing $v^u_0\underline{T_u}\wedge\lambda_{\pi_A}$, the effective direction simplifies to $V_0^u=-v_0^u$. Thus the Hamiltonian
\begin{equation}
	H_p=H_c+C_{\pi_\phi} + C_{\pi_a} + C_{A_\perp} + C_2,
\end{equation}
implements the constraints
\begin{subequations}\label{eq:constraints_list}
	\begin{align}
		\pi^j_{\phi}&=\epsilon^j{}_{bcd}\underline{e^b}\wedge\underline{\mathrm{D}}\phi^c\wedge\underline{e^d}+\underline{e^j}\wedge\underline{\mathrm{d}}\,\underline{A},\\
		\pi_{A}&=\eta_{ab}\underline{\mathrm{D}}\phi^a\wedge\underline{e^b},\\
		\underline{\mathrm{d}}\pi_{A}&=0,\\
		V_0^u\omega_\perp{}_{ua}\pi^a_{\phi}&=V_0^u\underline{\mathrm{D}}(\epsilon_{uabc}(\omega_\perp{}^a{}_i\phi^i\underline{e^b}\wedge\underline{e^c}
		+2e^a_\perp\underline{\mathrm{D}}\phi^b\wedge\underline{e^c})
		+\eta_{ua} e^a_\perp\underline{F}),
	\end{align}
\end{subequations}
and provides the consistency conditions
\begin{subequations}
\begin{align}
	\{C_{A_\perp},H_p\}&=\begin{multlined}[t]\int\underline{\mathrm{d}}\Big(-\eta_{ab}\omega_\perp{}^a{}_i\phi^i\underline{e^b}+\eta_{ab}\underline{\mathrm{D}}\phi^a e^b_\perp-\eta_{ij}\lambda^i_{\pi_\phi}\underline{e^j}\\
	{}-\underline{\mathrm{D}}(\lambda_2 V_0^u )\eta_{ua} e^a_\perp\Big)
	\wedge\underline{\mathrm{d}}A_\perp=0,\end{multlined}\\
	\{C_{\pi_A},H_p\}&\sim\begin{multlined}[t]
	\int\lambda_{\pi_A}\wedge\Big(\underline{T_u}(\lambda^u_{\pi_\phi}-\lambda_2 V_0^k \omega_\perp{}_{k}{}^u) \\
		{}+ \underline{\mathrm{D}}(\eta_{ab}\omega_\perp{}^a{}_i\phi^i\underline{e^b}-\eta_{ab}\underline{\mathrm{D}}\phi^a e^b_\perp-\eta_{ij}\lambda^i_{\pi_\phi}\underline{e^j}-\underline{d}(\underline{\mathrm{D}}(\lambda_2 V_0^u)\eta_{ua} e^a_\perp)\Big)=0,\end{multlined}\\
		\{C_{\pi_\phi},H_p\}&\sim\begin{aligned}[t]\int
		&\lambda^u_{\pi_\phi}\Big(-\epsilon_{umcd}\underline{e^c}\wedge\underline{e^d}\wedge\underline{\mathrm{D}}(\lambda_2 V_0^k \omega_\perp{}_{kn})\\
			{}-{}&\underline{\mathrm{D}}(\lambda_2 V_0^k )\wedge\epsilon_{kabc}\omega_\perp{}^a{}_u\underline{e^b}\wedge\underline{e^c}
		+\underline{T_u}\wedge\lambda_{\pi_A}\\
		{}-{}&\omega_\perp{}^a{}_u(\epsilon_{abcd}\underline{\mathrm{D}}\phi^b\wedge\underline{e^c}\wedge\underline{e^d}+\eta_{ab}\underline{e^b}\wedge\underline{F})\\
		{}+{}&\underline{\mathrm{D}}(\epsilon_{aucd}\omega_\perp{}^a{}_i\phi^i\underline{e^c}\wedge\underline{e^d}
		+2\epsilon_{aucd}e^a_\perp\underline{\mathrm{D}}\phi^c\wedge\underline{e^d}\\
		{}-{}&\eta_{ub} e^b_\perp\underline{\mathrm{d}}\,\underline{A}-\underline{\mathrm{D}}(\lambda_2 V_0^k )\wedge2\epsilon_{kauc}e^a_\perp\underline{e^c})\Big)=0,
		\end{aligned}\\
		\{C_2,H_p\}&\sim
		\begin{aligned}[t]
			\int&\lambda_2V_0^u\Big( \underline{\mathrm{D}}(\epsilon_{uabc}\omega_\perp{}^a{}_k\underline{e^b}\wedge\underline{e^c}\lambda^k_{\pi_\phi}
			- 2\epsilon_{uakc}e^a_\perp\underline{e^c}\wedge\underline{\mathrm{D}}\lambda^k_{\pi_\phi})\\
			{}+{}&
			\omega_\perp{}_{u}{}^k(-\epsilon_{abcd}\omega_\perp{}^a{}_k\underline{\mathrm{D}}\phi^b\wedge\underline{e^c}\wedge\underline{e^d}
-\eta_{ab}\omega_\perp{}^a{}_k\underline{e^b}\wedge\underline{\mathrm{d}}\,\underline{A}\\
		{}+{}&\underline{\mathrm{D}}(\begin{multlined}[t]\epsilon_{akcd}\omega_\perp{}^a{}_i\phi^i\underline{e^c}\wedge\underline{e^d}
		+2\epsilon_{akcd}e^a_\perp\underline{\mathrm{D}}\phi^c\wedge\underline{e^d}\\
		-\eta_{kb} e^b_\perp\underline{\mathrm{d}}\,\underline{A}
		+\lambda^i_{\pi_\phi}\epsilon_{ikcd}\underline{e^c}\wedge\underline{e^d}
		-\lambda_{\pi_A}\wedge\underline{e_k}
		))\end{multlined}
			\\
			{}+{}&\underline{\mathrm{D}}(\eta_{ua} e^a_\perp)\wedge\underline{d}\lambda_{\pi_A}\Big)=0,
			\end{aligned}
\end{align}
\end{subequations}
with $\sim$ meaning equality up to boundary equivalence. The result is a partial differential system. While an analytic method could mimic what was done for finding the secondary constraint, the full solution is far more complicated and actually unnecessary in order to understand the system, as well as the canonical behaviour of the theory. Equation $\{C_{A_\perp},H_p\}\equiv0$ is void, as it is a total derivative, while the rest form a linearly independent system, with eight equations for eight unknowns. No more constraints are found, when the background is well-behaving. This is, admittedly, a rather optimistic description, as the background dependence of the constraints is \emph{not} trivial, and can break at many parts.

After obtaining all of the functional derivatives of the constraints, it is possible to study the constraint algebra of $\{C_{A_\perp}, C_{\pi_A}, C_{\pi_\phi}, C_2\}$. The Gauss law constraint $A_\perp$ remains first class as in standard electromagnetism, as here it only has a nontrivial derivative w.r.t.\ $\pi_A$, so the Poisson brackets either vanish or turn out to be exact. The remaining three Poisson brackets are background dependent, but generally non-vanishing:
\begin{subequations}
	\begin{align}
		\{C_{\pi_A},C_{\pi_\phi}\}&\sim\int\underline{T_u}\wedge\lambda_{\pi_A}\lambda^u_{\pi_\phi},\\
		\{C_{\pi_A},C_{2}\}&\sim \int\lambda_2
		(\underline{\mathrm{D}}\lambda_{\pi_A}\wedge V_0^k(\omega_\perp{}_{k}{}^u\underline{e_u}-\underline{\mathrm{D}} e_{\perp k})
		-\lambda_{\pi_A}\wedge V_0^k\omega_\perp{}_{k}{}^u\underline{T_u}),\\
		\{C_{\pi_\phi},C_{2}\}&\sim
		\begin{multlined}[t]\int
		(
		-\underline{\mathrm{D}}(\lambda_2 V_0^k )\wedge\epsilon_{kabc}\omega_\perp{}^a{}_u\underline{e^b}\wedge\underline{e^c}
			+\underline{\mathrm{D}}(
			\lambda_2 V_0^a\omega_{\perp a}{}^k\epsilon_{kucd}\underline{e^c}\wedge\underline{e^d}\\
			-\lambda_2 V_0^k \underline{\mathrm{D}}(2\epsilon_{kucd}e^c_\perp\underline{e^d})))
		\lambda_{\pi_\phi}^u,\end{multlined}
	\end{align}
\end{subequations}
with the functional Lagrange multipliers to be taken as arbitrary in these expressions. So, the standard Gauss constraint is I class, while the secondary constraint and the definitions of momenta are II class. Altogether, the two electromagnetic polarizations are recovered, as the degree of freedom count provides $(14-2\cdot 1 - 3 - 4 - 1)/2=2$, background permitting. In this light, the isokhronon gauge theory coupled to khronon gravity might prove both more interesting and viable than the isokhronon on an arbitrary background alone, as the khronon provides torsionful solutions even in Minkowski, and the time gauge provides a natural direction for the secondary constraint.

Returning to the canonical structure, the analogy between the Lagrangian and Hamiltonian system is now plain. The isokhronon momentum becomes the spatial component of the integration constant, just as in the khronon theory~\cite{nikjoo2023hamiltonian}, while the longitudinal component $X_{\perp a}$ can be read off from the dynamics. As the Hamiltonian is first order and linear in momenta, the Lagrange multipliers $\lambda_{\pi_A}$ and $\lambda_{\pi_\phi}$ of the momenta correspond to the time evolution of the respective dynamical variables $\underline{A}$ and $\phi^a$. The integration constant 3-form $X_a$ is covariantly constant, as its canonical correspondent is simply the time evolution of the isokhronon momentum. Gauss's law is simply the spatial projection in~\eqref{eq:31_proto_inhom}, which is immediately picked up by the Lagrange multiplier $A_\perp$.

To finish this section, let us also show how the standard electromagnetic Hamiltonian can be recovered. Generally the purpose of constraints is to define the geometry of the phase space submanifold the theory inhabits. For practical purposes, they have to be solved explicitly. For most practical theories, the Lagrange multipliers are indeed simply algebraically fixed after solving the consistency conditions. The basic relations have to be established among the basic variables, here $\phi^a$, $\underline{A}$ and their momenta, but all of the obtained constraints~\eqref{eq:constraints_list} are principally differential equations, so this relation can be made proper only after introducing integration constants.

The simple descriptor of physical degrees of freedom is understood to be local and propagating, but a full understanding is more nuanced. Analytical solutions provide monopole, instanton etc. degrees of freedom. Some global degrees of freedom might be situated on the spacetime boundary and not affect physics in the bulk, but generally there is a great variety in their effects, cf.~\cite{Zlosnik:2018, Jirousek:2021, Jirousek:2022kli}. In a similar vein, a recent consideration of the requirements of quantum theory found that the Gauss law constraint could logically be relaxed, providing a ``shadow charge'' contribution with effects similar to cosmic dust or an electromagnetically charged environment~\cite{kaplan:2023a, kaplan:2023b}, cf. similar deliberation in~\cite{Golovnev:2023} and further background in~\cite{casadio2024relaxation}. Indeed, the entire vector-derivative approach to gauge theory and gravity has shown that such background phenomenology is possible already in classical theory, as a geometric and global effect. In the canonical description, the dynamics of momentum $\pi^a_\phi$, that is the Hamilton or Poisson equations of motion themselves, yield that $\pi^a_\phi$ is covariantly constant, thus essentially an integration constant, recovering~\eqref{eq:covariant_constant}.

It is possible to purely implicitly situate on the constraint surface, so that the constraints~\eqref{eq:constraints_list} are held and manipulated implicitly, without invoking an explicit parametrization of the constraint surface or a total Hamiltonian. Then after some manipulation of the constraints in 3+1 form the Hamiltonian on the constraint surface simplifies to
\begin{equation}\label{eq:basic_hamiltonian}
\begin{aligned}
	H_c&=\begin{aligned}[t]
		\int&\frac{1}{2}\underline{F}\wedge\underline{\star}\,\underline{F}
		+\frac{1}{2}\pi_A\wedge\underline{\star}\pi_A\\
		{}+{}&\underline{\mathrm{D}}^{\langle J}\phi^{I\rangle}\underline{\mathrm{D}}_{\langle I}\phi_{J \rangle}\underline{\text{Vol}}
		-(\underline{\star}\pi_{\phi I})\underline{e^I} \wedge\underline{F}\\
		{}-{}&\omega_\perp{}_{ai}\pi^a_\phi\phi^i,
	\end{aligned}
\end{aligned}
\end{equation}
where $\langle\rangle$ denote the symmetric trace-free component. The first line is simply Maxwell electromagnetism, while the second holds a vacuum null point energy shift and explicit magnetization, which is visible from the 3+1 split of $F\wedge H$ with a background excitation $H$. Note the conventional decomposition
\begin{equation}
	H=\mathrm{d}\sigma\wedge M - P
\end{equation}
for magnetization $M$ and polarization $P$. An effective polarization would couple to $\underline{\dot{A}}$, thus modifying the definition of momentum and Gauss law. However, magnetization and polarization are not Lorentz-invariant concepts, but will rather transform into each other in a covariant manner. The last line in the implicit Hamiltonian~\eqref{eq:basic_hamiltonian} is simply a manifold effect on velocity, and would disappear either on a flat background, when adopting a covariant definition of velocity which absorbs $\omega_\perp$, or when performing the analysis on a dynamical spacetime, rather than a static background. In principle, $\phi^a$ should be integrated out, preferring the electromagnetic and integration constant variables, but it would not elaborate the interpretation any further. Overall, the canonical description of the theory coincides with the Lagrangian understanding, although is notably more difficult to penetrate.

\section{Covariant phase space}
\label{sec:covariant}

\subsection{Covariant canonical geometry}
\label{subsec:covariant_geometry}

Recently there has been some wider interest into the covariant formulation of the phase space~\cite{Harlow:2019yfa, Freidel:2020xyx, Margalef:2021, ciambelli:2023, gieres2023covariant, assanioussi:2023, delValleVaroGarcia:2022ceq}. In this paper, the interest is in application of these methods to the isokhronon theory in~\cite{Gallagher:2022kvv} and elaborating on the recent~\cite{Gallagher:2023}. To be precise, there are several formulations of covariant canonical structure, see e.g. multisymplectic geometry, but this paper only considers what is known as the covariant phase space; these formulations are not completely independent either, see~\cite{gieres2023covariant} for more details. The covariant phase space can be seen as a dual canonical formulation of a physical theory versus the noncovariant Hamiltonian analysis, in the sense of establishing a symplectic structure on some phase space manifold (cf. the differential geometric formalization of classical mechanics). The formalism is implicitly covariant as it doesn't introduce the splitting of time and space, and Lorentz symmetry is implicitly preserved, as much as the solutions themselves do so\footnote{This is not always the case, even if the equations are written in a naively covariant way. Rarita-Schwinger higher spin theory~\cite{Rarita:1941mf} is a particular pathological example. To emphasize, the covariant phase space is as valid as, and defined to the extent of the solutions themselves, resp. Lagrangian dynamics.}, no observer is preferred by construction, and the phase space can be studied using geometric and algebraic methods. Note that the construction is \emph{not} in contention with the conventional phase space: see~\cite{Margalef-Bentabol_Villasenor:2022} for a recent proof of equivalence. Rather, it is a useful reformulation.

The basic construction of the covariant phase space considers the pre-phase space manifold $\tilde{\mathcal{P}}$ to be constituted from the trajectories of the system, that is the solutions to the equations of motion, and then builds a pre-symplectic form $\tilde{\Omega}$ on top. The initial configuration space consists of all possible field configurations $\mathcal{I}\equiv\Gamma(E)$, the (infinite dimensional!) space of sections $\Gamma(E)$ of the fiber bundle $E$ of whatever field theory is studied. The pre-phase space $\tilde{\mathcal{P}}$ is in essence a submanifold of this configuration space. The proper phase space $\mathcal{P}$ and symplectic structure $\Omega$ is obtained after factoring out degeneracies, respectively zero modes of $\tilde{\Omega}$, which are induced by (gauge) symmetries of the theory. In practice, the canonical structure is established through just a variation of the action, and the construction overall is heavily Lagrangian-based. For example, constraints and degrees of freedom are only implicitly defined by the solutions. At the same time, everything is constructed to be generally consistent and compatible with other formalizations of physics.

Let the theory be described by an action
\begin{equation}\label{eq:basic_action}
	S=\int_M L + \int_{\partial M}\ell,
\end{equation}
as in~\cite{Harlow:2019yfa}.
Let the boundary be $\partial M=\Gamma\cup\Sigma_+\cup\Sigma_-$, for spatial boundary $\Gamma$ and $\Sigma_\pm$ the future or past boundary. Generally the variation of the Lagrangian can be written as
\begin{equation}
	\delta L=\delta\phi^a\wedge E_a + \mathrm{d}\Theta,
\end{equation}
providing the Euler-Lagrange equations of motion $E_a=0$, and the pre-symplectic potential $\Theta$, used for defining the symplectic structure. In principle, for anticommuting, e.g. fermionic variables the left and right variations should be discerned. The trajectories of physically motivated examples are defined by a well-posed initial value problem and so are in correspondence with the Cauchy data of the system. Thus the basic assumption, once again, is that the manifold is globally hyperbolic, as it must admit at least some Cauchy surface; see again~\cite{oneill:1983,Wald:1984} for the causal structure details.

Defining the pre-symplectic form $\tilde{\Omega}$ requires interpreting the variation $\delta$ as an exterior derivative on the space of configurations, so differential forms are generally bi-graded on the bicomplex $\Omega^{p,q}(M\times\mathcal{I})$. The variation\footnote{In principle, the bi-grading provides a total exterior derivative $\bm{\mathrm{d}}=\mathrm{d}+\delta$ such that $\bm{\mathrm{d}}^2=\mathrm{d}^2=\delta^2=0$, so the compatibility of the differential in the graded algebra requires $\mathrm{d}$ and $\delta$ to anticommute, rather than commute. This is in contrast to~\cite{Harlow:2019yfa}, but for many practical cases this amounts to simply changing sign conventions. This analysis uses the commutative convention for simplicity, as it doesn't substantially change the canonical structure or the discussion.} $\delta$ then increases the vertical degree~$q$ (``across'' configurations), while the spacetime exterior derivative $\mathrm{d}$ increases only the horizontal degree~$p$ (``along'' spacetime).
As emphasized in~\cite{Harlow:2019yfa}, there can be a nontrivial exact ambiguity $\mathrm{d}C$ to the complete variation on the boundary
\begin{equation}
	B_\text{var}\equiv\Theta + \delta \ell - \mathrm{d}C.
\end{equation}
Then the pre-symplectic current reads as a pullback to $\tilde{\mathcal{P}}$ such that
\begin{equation}
	\omega\equiv\delta B_\text{var}\rvert_{\tilde{\mathcal{P}}}=\delta(\Theta-\mathrm{d}C)\rvert_{\tilde{\mathcal{P}}},
\end{equation}
and defines the pre-symplectic form
\begin{equation}
	\tilde{\Omega}=\int_\Sigma\omega.
\end{equation}
Note that $\tilde{\Omega}$ is independent of the choice of the Cauchy slice $\Sigma$. There can still be symmetry induced degeneracies, zero modes of $\tilde{\Omega}$, i.e.\ vector fields $\tilde{X},\tilde{Y}$ on $\tilde{\mathcal{P}}$ such that $\tilde{\Omega}(\tilde{X},\cdot)=\tilde{\Omega}(\tilde{Y},\cdot)=0$. Their Lie algebra closes, and they define a diffeomorphism subgroup $\tilde{G}$. The proper phase space factors this out, $\mathcal{P}=\tilde{\mathcal{P}}/\tilde{G}$. The induced symplectic form $\Omega$ is closed, unambiguous, non-degenerate and establishes a symplectic structure. Altogether, $\mathcal{P}$ behaves like what would be expected from a phase space, and there is a simple method to obtain the symplectic structure.

Separately, the canonical structure of topological theories is well-posed, see e.g.~\cite{gotay2004momentumI, gotay2004momentumII, Molgado_2019}, as is the case for metric theories. In the fancier pregeometric theories under study here, the problem lies in the change of the symmetry group structure of the background manifold, when transitioning from the metric to the topological regime. In particular, the orthonormal frame bundle loses meaning, as does the isometry group. It is possible to imagine a (inevitably discontinuous) transition form the metric-enforced $SO(1,3)$ subbundle to the $GL(4)$ principal bundle, and it is possible to talk about reductions of the structure group. Overall, however, this viewpoint remains to be explored in greater detail --- even if the basic quotient procedure remains unchanged.

Working in the space of fields, it is tempting to consider the field space itself with its symplectic structure as fundamental. In other words, spacetime geometry could be seen as the causal consequence of interacting materia. The question is how to make this notion precise, and consistent with observation. Many notions, e.g. geometric coincidence, are difficult to remove. There is some philosophical ambition if spacetime is not accepted as a separate entity\footnote{It is the dispute of substantivalism and relationalism of space and time, traced back most famously to Newton and Leibniz. Although perhaps somewhat formal or mainly philosophical, it still persists.}. If successful, quantum gravity, or a part of it, could arise naturally as derived from predisposed quantum fields. An initial step could be to derive causal structure from field structure, which would be interesting in its own right, but this remains to be done.

\subsection{Conserved currents}
\label{sec:conserved_currents}

Because the covariant phase space formalism operates through the variation, it is convenient to study conserved currents alongside the canonical structure. Do note, however, that generally conserved quantities do not necessarily have a symmetry behind them, that is the converse of Noether's theorem does not hold without additional assumptions, e.g. without instead working with quasisymmetries, see~\cite{Olver:1993}. For the theories under study, conserved currents were presented in~\cite{Gallagher:2023}, but this section also collects a few comments of general interest.

Let the Lagrangian be varied only up to a boundary term,
\begin{equation}
	\delta L=\mathrm{d}K.
\end{equation}
It can be argued from locality that necessarily $K=0$ for gauge parameters $\lambda_\alpha$ with compact support~\cite{Avery:2015rga}; the same holds for exact symmetries. In the differential geometry of the covariant phase space this variation is rather thought as a symmetry vector field acting on the general variation, $\delta L\to X\cdot\delta L$, but the simpler functional counterpart keeps the transformation implicit in the field variations $\delta\phi^a$ in the expansion of $\delta L$, see also e.g.~\cite{Mitsou:2014}. Regardless of this, Noether's theorems, including the first, the second, and the boundary theorem, proceed by separating variations in the bulk and on the boundary, comparing on- and off-shell behaviour, and separating gauge transformation parameters --- see~\cite{Brading2003-BRASAN-2} for much more details.

The commonly reported Noether's first theorem is applied only for global active symmetries of the action, where the derived conserved currents and charges are generally nontrivial and physically meaningful. However, the derivation holds for local, including gauge, symmetries just as well, and still provides a conserved current $J$ such that $\mathrm{d}J=0$. It is important to emphasize that the nature of symmetry transformations can inherently differ. Gauge transformations in particular do not represent an inherent symmetry of the system, but rather represent redundancy of the description. Thus, conventionally, the resulting currents for gauge symmetries are considered to be unphysical due to several issues: they explicitly depend on a transformation parameter, which cannot generally be factored out, thus are not thought to be strictly gauge invariant and instead define a family of currents, rather than a single observable with a clear physical meaning. They do not necessarily represent useful observables, as per~\cite{Karatas_Kowalski:1989} ``...they fail in what is the signal function of a conserved charge, namely, to provide a labeling of equivalence classes of systems governed by the same underlying dynamics,'' and the gauge symmetry must ultimately be broken to solve the equations of motion uniquely.

Therefore, Noether's first theorem is primarily of interest for global symmetries, or at most in a mixture of global and local symmetries~\cite{Aoki:2022}, while it is the somewhat less known Noether's second theorem that is applicable for infinite-dimensional Lie groups and local symmetries. The second theorem yields constraints which must hold off-shell, see~\cite{Avery:2015rga, Kosyakov:2007}. Here it is more convenient to follow a more differential-geometric description, adapting~\cite{Brading_Brown:2000, Brading2003-BRASAN-2} similarly to~\cite{Haro:2022}. So, assuming $L=L[\phi^a,\mathrm{d}\phi^a]$ for~\eqref{eq:basic_action}, the variation provides exactly the standard results for Noether currents:
\begin{equation}\label{eq:Noether_derivation}
	\begin{aligned}
		\delta S[\phi^a,\mathrm{d}\phi^a] &=\int_M (\delta\phi^a \wedge E_a + \mathrm{d}\Theta)  +\oint_{\partial M}\delta \ell=\\
		&=\int_M \bigg[\delta \phi^a\wedge\frac{\partial L}{\partial \phi^a} + \mathrm{d}\delta\phi^a\wedge\frac{\partial L}{\partial\mathrm{d}\phi^a}\bigg]   +\oint_{\partial M}\delta \ell=\\
		&=\int_M \bigg[\delta \phi^a\wedge\bigg(\frac{\partial L}{\partial \phi^a}-(-1)^p\mathrm{d}\frac{\partial L}{\partial\mathrm{d}\phi^a}\bigg)
		+ \mathrm{d}\bigg(\delta\phi^a\wedge\frac{\partial L}{\partial\mathrm{d}\phi^a}\bigg)\bigg]+\oint_{\partial M}\delta \ell\equiv\\
		&\equiv \int\mathrm{d}K.
	\end{aligned}
\end{equation}
Restricting to classical solutions provides the Noether current
\begin{equation}
	J=\Theta - K,
\end{equation}
which satisfies the continuity equation $\mathrm{d}J=0$, and integrated over some 3-dimensional volume $V$ defines the Noether charge
\begin{equation}
	Q=\int_V J.
\end{equation}
Furthermore, by the Poincaré lemma, there is a superpotential $U$ such that
\begin{equation}
	J=\mathrm{d}U,
\end{equation}
topology permitting.

Since the transformations are arbitrary, possibly off-shell, it is likewise possible to choose such $\delta\phi^a$ which vanish on the boundary, so the contribution
\begin{equation}
	\int_{\partial M} (\Theta + \delta\ell)=\int_{\partial M}\bigg(\delta\phi^a\wedge\frac{\partial L}{\partial\mathrm{d}\phi^a} + \delta\ell\bigg)=0,
\end{equation}
leaving separately
\begin{equation}
	\delta \phi^a\wedge\bigg(\frac{\partial L}{\partial \phi^a}-(-1)^p\mathrm{d}\frac{\partial L}{\partial\mathrm{d}\phi^a}\bigg)=\delta \phi^a\wedge E_a=0.
\end{equation}
This can be used to define constraints on the symmetry transformations themselves. Now conventionally the infinitesimal variation $\delta\phi^a$ is only considered to first order in derivatives of the gauge parameters $\lambda_\alpha$, i.e.\ only the first terms in a series like
\begin{equation}
	\delta\phi^a=\sum_\alpha \Big(a^a_{\alpha}(\phi^a,\partial_\mu\phi^a,x^\mu)\delta\lambda_\alpha(x^\mu) + b^{a\mu}_{\alpha}(\phi^a,\partial_\mu\phi^a,x^\mu)\partial_\mu(\delta\lambda_\alpha(x^\mu))\Big),
\end{equation}
although this can be generalized to arbitrary order as in the original publication~\cite{Noether:1918}. However, the theorem's procedure only requires to separate the bulk and the boundary contributions, so it is attractive to \emph{define} the symmetry transformation using two arbitrary parameters $a^a_\alpha$ and $b^a_\alpha$ as
\begin{equation}\label{eq:Noether_series_exp}
	\delta\phi^a(\lambda_\alpha, \mathrm{d}\lambda_\alpha)=\sum_\alpha \Big(\delta\lambda_\alpha\wedge a^a_{\alpha} + \mathrm{d}\delta\lambda_\alpha\wedge b^a_{\alpha}\Big).
\end{equation}
This is not necessarily the most general resummation formula, but a convenient parametrization. There is, however, merit in such a simple sum, cf. Hodge decomposition of a form into closed, exact, and harmonic components on a closed Riemannian manifold.

A simple shuffling under the integral sign using Stokes' theorem yields
\begin{equation}
	\sum_\alpha \delta\lambda_\alpha\wedge a^a_{\alpha}\wedge E_a - \sum_\alpha(-1)^k\delta\lambda_\alpha\wedge\mathrm{d}(b^a_{\alpha}\wedge E_a)=0,
\end{equation}
which for otherwise arbitrary $\delta\lambda_\alpha$ implies
\begin{equation}
	a^a_{\alpha}\wedge E_a = (-1)^k\mathrm{d}(b^a_{\alpha}\wedge E_a).
\end{equation}
This is a strong equation and is independent of the gauge parameters. Carefully keeping track of the interior and boundary also yields the boundary theorem --- by similar arguments the boundary contribution has to vanish independently, see~\cite{Brading2003-BRASAN-2}. A classical example of applying Noether's second theorem would be deriving the electromagnetic 4-current, and this would yield that the matter field equations are a sufficient, but not a necessary condition for deriving the continuity equation~\cite{Brading_Brown:2000}.

What is important is that Noether's theorem does not necessarily provide conserved currents if background fields also take part in the symmetry transformations, even if in total the action is invariant, see also~\cite{Brading_Brown:2000, Banados_Reyes:2016}. Indeed, such deficiency of conservation can be seen in the first variation~\eqref{eq:Noether_derivation}: for any background field $\phi^a$ with a symmetry variation $\delta\phi^a$, its corresponding Euler-Lagrange term does not necessarily vanish on-shell, regardless of the total variation. Contrariwise, if the background field does not take part of the symmetry transformation, there might be no symmetry at all. In principle, it is possible for the Euler-Lagrange term to still vanish for specific field configurations which are fixed or constrained compatibly, but this does not hold generically. It is possible to further study conserved quantities on such ``misaligned'' backgrounds, although this doesn't appear to be common, and the first variation already reveals a significant portion of such analysis, as the remaining non-vanishing terms quantify the lack of conservation of the current prototype.

\subsection{Quantization}
\label{subsec:quantization}

An intriguing application of symplectic structure is quantization. The contemporary understanding of quantum physics ranges from quantum information theory to Feynman path integrals to (perturbative) algebraic quantum field theory to lattice calculations and beyond, and the calculations of practically and phenomenologically useful quantities such as correlators and Feynman rules are left to a future study. Nevertheless, as the symplectic structure relates to the Poisson structure, and thus to canonical quantization, it is possible to derive some interesting results from canonical geometric (and algebraic) considerations alone. This section briefly describe three approaches: geometric quantization, deformation quantization, and the BRST-BV formalism. The main point is to emphasize that the symplectic structure can be seen as the natural arena for quantization.

Canonical quantization procedures generally follow Dirac's quantum conditions for the observable algebra. So, in particular, the mapping $f\mapsto\hat{f}$ from classical observables $f\in C^\infty(M)$ to quantum operators $\hat{f}\in\mathcal{O}$ should be
\begin{enumerate}
	\item linear,
	\item give $\hat{f}$ as the multiplication operator when $f=\mathrm{const}$,
	\item $\{f_1,f_2\}=f_3\Rightarrow[\hat{f}_1,\hat{f}_2]=-i\hbar\hat{f}_3$.
\end{enumerate}
Sometimes the completeness condition is added, but generally these conditions alone do not provide a satisfactory quantization map --- see discussion on the Gronewold-van Hove theorem~\cite{Gotay:1999}. Further subtleties abound, especially for gauge theory where the mathematical description of the physical degrees of freedom is extraneous, requiring gauge fixing; or, how in non-Abelian gauge theory renormalization and anomaly cancellation necessitates more sophisticated ghost- and anti-field schemes. Simultaneously the quantization schemes strive to be consistent with each other, while the differences in philosophy \emph{and} construction can be quite substantial. But what is emphasized here is that the covariant phase space is one definite and (relatively) well defined way to describe the arena where all physics, including \emph{quantum} physics, should take place in.

So, the symplectic structure quickly directs to the formalism of geometric quantization, suggested to perhaps even assist in the corner algebra approach to quantum gravity~\cite{ciambelli:2023}. Although well-motivated, its simplicity proves to be deceptive, as the method itself quickly runs into mathematical difficulties, and outside of the famous results of~\cite{Witten:1988hf} and the like, explicit application to field theory in standard spacetime appears to be rather scarce\footnote{As to spinor or even more so to interacting fields. But the complete rigorous development of interacting quantum field theory remains an unsolved problem, so it is not a issue unique to geometric quantization.}. Generally, accounts on the applicability to field theory are rather contradictory, although e.g. deriving the Fock space of the free scalar field is a worked out example, see~\cite{Woodhouse:1991,Nair:2005}, also~\cite{Helein:2012}, and discussion of geometric quantization of fermions in~\cite{Kostant:1977}. But the field theory suggestion remains to follow the symplectic structure of the covariant phase space, or equivalent.

The standard reference is~\cite{Woodhouse:1991} and a recent introduction in~\cite{Berman_Cardoso:2022}, but the formalism has seeped into wider mathematical quantum mechanics literature, as the basic procedure is rigorous. Very briefly, beginning with a symplectic manifold $(M,\omega)$, the aim is to explicitly construct the system's Hilbert space $\mathcal{H}$ and realize Dirac's quantum conditions for the observable algebra $\mathcal{O}$. Skipping a series of ansatz and geometric discussion, the solution of geometric quantization is to realize the quantization mapping via the Kostant-Souriau (pre)quantum operator. It appears to explicitly depend on the symplectic potential $\theta$, while taking a form akin to a covariant derivative. Therefore, the geometrically natural resolution is to build a Hermitian line bundle $\pi:B\to M$ on the symplectic manifold. The symplectic potential is promoted to a connection, the curvature proportional to the symplectic form, and a pre-Hilbert space $\mathcal{H}_\text{pre}$ consisting of square integrable sections $s:M \to B$. All this is possible only when the symplectic form satisfy certain (Weil) integrality conditions --- in particular, simply connected manifolds behave well.

But the prequantization construction alone is not satisfactory. For one, the Hilbert space is too big, admitting wave functions that depend on \emph{both} position and momentum, requiring a choice of polarization, if even possible. The inner product may diverge, requiring half-form quantization. Further still, a Hamiltonian vector field might depolarize sections under its flow, e.g. time evolution, necessitating the Blattner-Kostand-Sternberg construction --- this all without touching field theory, where the phase space is infinite-dimensional.

Deformation quantization seems to have enjoyed some greater success as a formalization of quantum field theory than geometric quantization, as it allows for a simpler relation to perturbative calculations, and the structures are easier defined, see~\cite{Hirshfeld:2002, Galaviz_et_al:2008, Hollands:2008, Waldmann:2016, Hawkins_Rejzner:2020}. It is not in opposition to geometric quantization, however, as it still bases on either the symplectic~\cite{Fedosov:1994} or Poisson~\cite{Kontsevich:2003} structures. The issue of non-perturbative treatment of quantization remains, resp. the distinction between formal and strict deformation quantization, and existence or convergence proofs require attention.

In a few words, beginning with a Poisson manifold $(M,\pi)$, the idea is to \emph{deform} the associative algebra of classical observables, i.e.\ smooth real-valued functions under the pointwise multiplication, to a noncommutative associative algebra $\mathcal{A}$ of complex functions under a star product $\star$, which satisfies certain (physical) consistency conditions. For field theory, the Poisson bracket can be realized through the Peierls bracket, which in turn is consistent with the symplectic structure of the covariant phase space, and classical and quantum observables are complex-valued functionals on the space of (smooth) sections. The $\star$-product is understood as a bilinear mapping to the formal power series $\mathcal{A}\times\mathcal{A}\to\mathcal{A}[[\hbar]]$,
\begin{equation}
	F\star G=\sum_{n=0}^\infty \hbar^n C_n(F,G),
\end{equation}
where $C_n$ are bidifferential operators that vanish on constants, and satisfy certain (physical) conditions --- in particular, the classical limit $C_0(F,G)=FG$ and the correspondence principle $C_1(F,G)-C_1(G,F)=i\hbar\{F,G\}$. The interacting theory of the action $S=S_0+\lambda V$ can follow from the retarded Møller operator, as it intertwines the Peierls bracket~\cite{Dito:1990rj,Dito:1993yv}, or the interacting fields can be constructed using the formal S-matrix and the quantum Møller operator~\cite{Hawkins_Rejzner:2020}.

Finally, particularly relevant to (especially non-Abelian) gauge theory is the BRST or BV formalism, see~\cite{Henneaux:1992ig} for a detailed or~\cite{Henneaux:1989jq, Fuster:2005eg} for a shorter introduction. The BRST formalism provides a consistent prescription for introducing (Grassmann-graded) ghosts-antifields and gauge fixing in the quantum gauge theory. For this purpose, the original gauge invariance is replaced by the BRST global symmetry $\mathfrak{s}$, retained even after gauge fixing, which acts on the extended phase space including the additional non-physical fields. The purpose is for BRST invariance to be a substitute for gauge invariance. It is nilpotent, $\mathfrak{s}^2=0$, and as such it allows for a cohomological interpretation, so that of particular importance the zeroth cohomology group $H^0(\mathfrak{s})=C^\infty(\tilde{\mathcal{P}}/\tilde{G})$ holds the gauge invariant functions, i.e.\ observables --- functions on the covariant phase space. The additional fields define an antibracket $(\cdot,\cdot)$ structure, so the master equation $(S,S)=0$ allows for solving for the generalized action $S$, i.e. the symmetry generating function in $\mathfrak{s}\Phi=(S,\Phi)$.

The proof and justification of the BRST procedure is significantly lengthier. What is noteworthy is that the BRST symmetry realizes the gauge reduction $\mathcal{I}\to\tilde{\mathcal{P}}\to\tilde{\mathcal{P}}/\tilde{G}=\mathcal{P}$ from the space of all configurations to the space of solutions to the space of solutions modulo gauge symmetry, i.e.\ the covariant phase space, through two nilpotent operators. These operators constitute (a part of) the BRST symmetry transformation $\mathfrak{s}$~\cite{Henneaux:1989jq}. The reduction to the space of solutions $\tilde{\mathcal{P}}$ is realized by the Koszul-Tate resolution and the differential $\delta$. Reducing gauge symmetry to $\tilde{\mathcal{P}}/\tilde{G}$ is realized by another differential $\mathrm{d}$, the vertical exterior derivative along the gauge orbits. Altogether, the BRST symmetry transformation is $\mathfrak{s}=\delta+\mathrm{d}+\text{extra}$, as is often written, with $\mathfrak{s}^2=0$ ensured by the extra terms of higher antighost number, determined by a more precise action on the fields. The BV formalism is similar, but with larger applicability, viz. open and reducible algebras, although the exact limits between the BRST and BV formalism are not necessarily precisely delineated. The quantum theory is obtained by writing the path integral, either in the Hamiltonian or Lagrangian variant. The latter requires a few extra steps for gauge fixing.

\subsection{First variation}
\label{subsec:variation}

The preceding lengthy preambula serves the purposes of the basic procedure of the first variation. The conserved currents were already studied in~\cite{Gallagher:2023}, and a lengthy addendum to them was presented here in section~\ref{sec:conserved_currents}, so the focus is on what other conclusions can be derived from the symplectic structure. The three electromagnetic theories of interest are, up to (coupling) constants,
\begin{align}
	S_\text{EM}&=\int F\wedge *F,\label{eq:action_EM}\\
	S_{\text{EM}^1}&=\int\frac{1}{2}B\wedge*B+B\wedge F,\label{eq:action_EM1st}\\
	S_{\phi}&=\int\frac{1}{2}\epsilon_{abcd} \mathrm{D}\phi^a\wedge e^b\wedge \mathrm{D}\phi^c\wedge e^d + \eta_{ab} \mathrm{D}\phi^a\wedge e^b\wedge F,\label{eq:action_isokhronon}
\end{align}
respectively standard Maxwell electromagnetism, its first order variant, and the recent isokhronon electromagnetism of primary interest in this paper. Here the Hodge star being $*$. Alternatively, the first order action can be written through $\tilde B=*B$, but the current form has the benefit of making the BF-theory relation more apparent. This analogy can be pushed further, see e.g.~\cite{Cattaneo_et_al:1998}.

Deriving the total variation is a small exercise in the variational calculus of forms, only the Hodge star offers subtlety. With a more in-depth explanation in~\cite{Itin:2023}, there are three interdependent variations for a differential $k$-form $\mu=\frac{1}{k!}\mu_{i_1\ldots i_k} e^{i_1}\wedge\ldots\wedge e^{i_k}$, related by
\begin{equation}\label{eq:dependent_variations}
	\delta \mu=\frac{1}{k!}\delta \mu_{i_1\ldots i_k} e^{i_1}\wedge\ldots\wedge e^{i_k} + \delta e^a\wedge\tetrad_a\lrcorner\mu,
\end{equation}
Only two are independent, most conveniently the differential form $\mu$ and the coframe $e^a$. Now it is possible to derive a relation for the variation of the Hodge star
\begin{equation}\label{eq:Hodge_star_variation}
	\delta*\mu=*\delta\mu - *(\delta e^a\wedge \tetrad_a\lrcorner \mu) + \delta e^a\wedge\tetrad_a\lrcorner*\mu,
\end{equation}
which crucially mediates the gravitational and curvature effects on the physical fields in most standard actions. Alternatively it might be simpler to follow the variation just in the components per~\eqref{eq:dependent_variations}.

Altogether
\begin{align}
	\delta S_\text{EM}&=\begin{multlined}[t]\int2\delta A\wedge\mathrm{d}*F +2\mathrm{d}(\delta A\wedge*F)\\
	{}+ \delta e^a\wedge( - (\tetrad_a\lrcorner F)\wedge *F +  F\wedge \tetrad_a\lrcorner*F),\end{multlined}\\
	\delta S_{\text{EM}^1}&=\begin{multlined}[t]\int\bigg(\delta B\wedge(*B + F) -\delta A\wedge \mathrm{d}B + \mathrm{d}(\delta A\wedge B)\\
		{}+\delta e^a\wedge\frac{1}{2}(-(\tetrad_a\lrcorner B)\wedge* B + B\wedge \tetrad_a\lrcorner* B),\end{multlined}\\
	\delta S_{\phi}&=\begin{aligned}[t]
	\int\bigg(&\delta e^a\wedge (\epsilon_{abcd} \mathrm{D}\phi^b\wedge \mathrm{D}\phi^c\wedge e^d -  \eta_{ab} \mathrm{D}\phi^b\wedge  F)\\
		{}+{}&\delta\omega^a{}_i \wedge (\phi^i(\epsilon_{abcd}  e^b\wedge \mathrm{D}\phi^c\wedge e^d + \eta_{ab} e^b\wedge F))\\
		{}+{}&\delta A\wedge\mathrm{D}(\eta_{ab} \mathrm{D}\phi^a\wedge e^b)\\
		{}+{}& \mathrm{d}(\delta A\wedge\eta_{ab} \mathrm{D}\phi^a\wedge e^b)\\
		{}-{}&\delta\phi^a\mathrm{D}(\epsilon_{abcd}  e^b\wedge \mathrm{D}\phi^c\wedge e^d + \eta_{ab} e^b\wedge F)\\
		{}+{}&\mathrm{d}(\delta\phi^a(\epsilon_{abcd}  e^b\wedge \mathrm{D}\phi^c\wedge e^d + \eta_{ab} e^b\wedge F))\bigg).
	\end{aligned}
\end{align}
These variations also include the gravitational-geometric coframe and spin connection fields, although they are relatively inconsequential without a theory of gravitational dynamics attached, and should be entirely removed when taken as a background. The coframe variations provide the proper energy-momentum 3-form
\begin{equation}
	\theta_a=- \tetrad_a\lrcorner F\wedge *F +  F\wedge \tetrad_a\lrcorner*F,
\end{equation}
in components (or, dual tensor densities)
\begin{equation}
	\theta_a{}^k\sim F_{ai}F^{ki} - \frac{1}{4}F_{ij}F^{ij}\delta_{a}^{k}.
\end{equation}
The first order variants require going on-shell, and the isokhronon provides an effective integration constant background contribution. The spin connection variation in $S_{\phi}$ is strict to the point of breaking the theory, but can be thought to imply that the $\phi^a$ theory requires a compatible theory of gravity (or, a choice of background) to be viable. The dynamical consequences for the rest of the variations is just to enforce the Dirichlet boundary conditions, as in~\cite{Harlow:2019yfa}.

The symplectic structure itself is more interesting. The pre-symplectic forms read
\begin{align}
	\tilde{\Omega}_\text{EM}&=\int_\Sigma\delta A\wedge *\delta F,\\
	\tilde{\Omega}_{\text{EM}^1}&=\int_\Sigma\delta A\wedge \delta B,\\
	\tilde{\Omega}_{\phi^a}&=\int_\Sigma\delta A\wedge  \delta(\eta_{ab}\mathrm{D}\phi^a\wedge e^b),
\end{align}
where global constants and background variations were removed. The symplectic form is defined on the covariant phase space, therefore restricted to solutions of the equations of motion, which provide
\begin{align}
	B&=F,\\
	\eta_{ab}\mathrm{D}\phi^a\wedge e^b&=*F+\frac{1}{2}(*X_a)\wedge e^a,\,\mathrm{D}X_a=0.
\end{align}
As $\delta$ can be thought of not as an independent variation, but an exterior derivative in the space of functionals, the symplectic structure of first order electromagnetism coincides with the second order form, while the $\phi^a$ theory introduces an integration constant modification.

Standard first order electromagnetism and Yang-Mills theory is indiscernible from the standard second order Maxwell form. This is obvious in the classical level, has been known to be true on the quantum level, and the Yang-Mills case was recently studied in the covariant BV formalism~\cite{Lavrov:2021} (see~\cite{Cattaneo_et_al:1998} for an earlier discussion in the same vein). However, from the earlier discussion on quantization, it now appears that the covariant phase space and the canonical quantization procedure implies quantum equivalence to be almost trivial, for both electromagnetism and non-Abelian theory. The auxiliary field is entirely integrated out, and the symplectic structures entirely agree. At most the auxiliary $B$-field remains in a direct product structure, but taking the equality as identification would dismiss even this. As the BRST-BV formalism still ultimately probes the covariant phase space, an equality (isomorphism) of the covariant phase spaces suggests quantum equivalence even for non-Abelian theory. This deserves emphasis, that studying the canonical symplectic structure of the theory derives some far-reaching conclusions, classical \emph{and} quantum, using rather simple methods. 

It is possible to characterize the quantum theory of the isokhronon $\phi^a$. Any possible difference comes from the torsion-mediated integration constant background $\frac{1}{2}(*X_a)\wedge e^a$, which unfortunately lacks a simple interpretation. Nevertheless, it is part of the description of the solutions. The non-covariant analysis shows that $X_a$ does not define any propagating degrees of freedom. Instead, the integration constants define the embedding of the stationary surfaces. In restricting to the covariant phase space, the only remaining configurations are the field solutions of the gauge field, which are simply deformed by an addition of the integration constant background, \emph{including} in the quantum theory. Therefore the \emph{entire} phenomenology of the $\phi^a$ theory is that of gauge theory on a classical background, both classically and quantum-wise. Refer to the excellent overview~\cite{Fedotov:2023} for reference how to describe quantum fields on a classical background. In practice for the theory in question, the phenomenology is understood completely if the background is understood completely.

Still, it is possible to understand the space of quantum states. The covariant phase space, in its standard formulation, includes \emph{all} solutions of the equations of motion. The trajectories of the $\phi^a$ theory require extra data to be defined, that is the $X_a$ electromagnetic background, but in principle all possible backgrounds satisfying $D X_a=0$ define additional solutions which inhabit the phase space. Therefore, when the covariant phase space is taken as the canonical structure upon which to build a quantum theory, the Hilbert space becomes awkwardly large. Furthermore, the states of different backgrounds are not necessarily Dirac delta-orthogonal to each other, permitting quantum transitions. This is easiest to argue in non-relativistic quantum mechanics, where the inner product between two wave functions of two different backgrounds (say, minorly perturbed with respect to each other) is not necessarily orthogonal. In relativistic theory this requires handling asymptotic Fock states\footnote{For reference, this would further lead into the complicated discussion of Haag's theorem~\cite{Haag:1955ev}, see also~\cite{Maiezza:2020qib}, which we will not currently pursue.}, as the Hilbert space of fields undergoing interaction is, generally speaking, not entirely understood outside special cases.

This provides a similar quandary to argue for the shadow charge approach~\cite{kaplan:2023a, kaplan:2023b}. The natural extension is to a \emph{total} Hilbert space of all possible backgrounds. It is not immediately clear why the states should be orthogonal. It could be argued that the selection of physically available observables themselves for any single experiment is limited by the background in which the measurement takes place. However, this does not appear a strong counterpoint as all of the other observables still in principle exist. In other words, it would require introducing an additional concept of a physical Hilbert space, which has been suitably restricted.

These results can be understood as suggesting that background independence is not just a requirement of general relativity, but of any physical theory. More explicitly, there is no immediate restriction on adding an arbitrary background term
\begin{equation}
	S_J=\int A\wedge J,
\end{equation}
if the background current $J$ is conserved, cf. transforms per Noether's theorems. It would not enforce any dynamical consequences outside of the background current. Yet such backgrounds have not been observed on a fundamental level, but only used in effective theory, as in e.g. a high intensity laser field. A more nuanced interpretation would suggest that the mathematical formalism should be adapted to forbid such terms --- perhaps by integrating all possible currents out, taken as a definition of the theory, or by using some other formalism where such ambiguity does not come up. Similarly, the thermodynamical implications can provide additional clarity. The minimal solution is, of course, to simply never introduce the background terms to the action.

\section{Duality}
\label{sec:duality}

There are several points to add about the duality of gauge theories. Indeed, the first order formalism, where the constitutive law is made dynamical, also precisely provides a parent action used in deriving dual theories. Electromagnetism is famously self-dual in this sense, while the case for Proca to Kalb-Ramond, and Maxwell-Chern-Simons, as will be studied, is more subtle. The $S_\phi$ theory itself, and other similar modifications, are thus in a curious class of modified parent actions with an electromagnetic limit. First, it can be seen that the quantum implications of the covariant symplectic apparatus appear in agreement with the path integral approach. Further, the Proca-Kalb-Ramond duality is, once again, emphasized, but in particular it is shown that quartic modifications are no longer dual. Thus, their limits need not agree, in opposition to~\cite{Hell:2022}. The Hamiltonian formalism is immediately consistent with dimensional reduction. Notably, this leads to show that the Maxwell-Chern-Simons duality~\cite{Armoni:2022xhy} also follows from the standard duality integration procedure, provided the vacuum energy is properly handled. Proca-Chern-Simons becomes a direct continuation. Finally, the differential geometry of duality gives a simple explanation to duality \emph{rotations}, especially to the issue that non-Abelian Yang-Mills theories lack such transformations, i.e.\ gives a global description to~\cite{Deser:1976iy}.

\subsection{Path integral}
\label{subsec:path_integral}

Even though some technicalities of the path integral remain somewhat contentious, e.g. in the measure theory on the field space, the physical analysis is well established. Then, gauge theory requires a suitable gauge fixing scheme, e.g. the BRST complex as discussed before. The premise here remains the same as in standard first order theory~\cite{Lavrov:2021}.

To the basic fields $A$ and $\phi^a$, the BRST formalism introduces the ghost $C$, antighost $\bar{C}$ and auxiliary (gauge fixing) $b$ fields, in addition to their antifields $C^*$, $\bar{C}^*$ and $b^*$. In the electromagnetic case, the vector potential transforms according to $A\to A + \mathrm{d}\varphi$, while the isokhronon transforms trivially as $\phi^a\to\phi^a$. Therefore, the solution to the BRST master equation is essentially unchanged from electromagnetism, cf.~\cite{Henneaux:1992ig},
\begin{equation}\label{eq:phi_BRST_solution}
	S_{\phi,\text{BRST}}=S_\phi + \int A^*\wedge *\mathrm{d}C + i \int \bar{C}^*\wedge*b.
\end{equation}
Continuing further, various gauge fixing fermions could be considered, implementing the Faddeev-Popov action etc., but the crucial point is that the isokhronon auxiliary vector, just as standard first order theory, does not change the gauge fixing component of the path integral.

Thus it is sufficient to only consider the component of the generating functional that has the definition of the basic gauge theory, i.e.\ the path integral without the gauge fixing contribution,
\begin{equation}\label{eq:isokhronon_parent_action}
	Z_\phi=\int\mathcal{D}A\ \mathcal{D}\phi^a\exp\bigg[i\int\frac{1}{2}\epsilon_{abcd}\mathrm{D}\phi^a\wedge e^b\wedge\mathrm{D}\phi^c\wedge e^d + \eta_{ab}\mathrm{D}\phi^a\wedge e^b\wedge F\bigg].
\end{equation}
Similarly, the external sources $J_A$ and $J_\phi$ in the generating functional can be put aside, especially as for vacuum functionals $J_A=J_\phi=0$. Integrating out $\phi^a$ characterizes the quantum phenomenology. This is simplest done after a field shift
\begin{equation}
	\phi^a\to\Phi^a + f^a,
\end{equation}
where $\Phi^a$ is a classical solution to the equations of motion, so
\begin{align}
	\epsilon_{abcd}e^b\wedge\mathrm{D}\Phi^c\wedge e^d + \eta_{ab}e^b\wedge F&=X_a,\mathrm{D}X_a=0,\\
	\mathrm{D}\Phi^a\wedge e^a&=*F + \frac{1}{2}(*X_a)\wedge e^a.
\end{align}
As $A\to A$ remains unchanged, the functional Jacobian is trivial, and the action decouples entirely:
\begin{equation}
	Z_\phi\to\int\mathcal{D}A\ \mathcal{D}f^a\exp\bigg[i\int\frac{1}{2}*F\wedge F + \frac{1}{4}(*X_a)\wedge e^a\wedge F + \epsilon_{abcd}\mathrm{D}f^a\wedge e^b\wedge\mathrm{D}f^c\wedge e^d\bigg].
\end{equation}
This produces the integration constant background for standard electromagnetic theory also in the quantum regime, and verifies the symplectic description provided earlier. Note the covariant phase space approach only required a geometric or symplectic understanding of the phase space to also describe its quantum behaviour.

The decoupled massless vector theory $f^a$ could be integrated out and removed, dismissing the decoupled propagator, but as later seen in the Maxwell-Chern-Simons case this is generally not allowed. Duality should not actually change the master system or path integral, only the variables used. Integrating out $A$ instead produces the dual magnetic theory, as the resulting Dirac delta enforces
\begin{equation}
	\delta(\mathrm{d}(\eta_{ab}\mathrm{D}\phi^a\wedge e^b))\Rightarrow \eta_{ab}\mathrm{D}\phi^a\wedge e^b=\mathrm{d}a
\end{equation}
for some new (``magnetic'') potential $a$. Note that $f^a$ compares as
\begin{equation}
	\int\epsilon_{abcd}\mathrm{D}f^a\wedge e^b\wedge\mathrm{D}f^c\wedge e^d\sim\int\mathrm{D}\tetrad_a\lrcorner\phi\wedge\mathrm{D}\tetrad_b\lrcorner\phi\wedge\star(e^a\wedge e^b).
\end{equation}
to a usual massless vector action
\begin{equation}
	\int \mathrm{d}A\wedge*\mathrm{d}A\sim\int\tetrad_a\lrcorner\mathrm{d}A\wedge\tetrad_b\lrcorner\mathrm{d}A\wedge\star(e^a\wedge e^b),
\end{equation}
with the difference being in various Lie derivative terms. The Hodge star $*$ and Lorentz dualization $\star$ agree on a surface basis $e^a\wedge e^b$, to bypass any explicit metric inverses and as motivated by Lorentz gauge theory~\cite{Zlosnik:2018}. However, phenomenology-wise the isokhronon $\phi^a$ can be seen as electromagnetism with a background $\frac{1}{2}(*X_a)\wedge e^a, \mathrm{D}X_a=0$.

\subsection{Proca and Kalb-Ramond theory}
\label{subsec:Proca_Kalb_Ramond}

A direct follow-up to Maxwell theory is Proca theory. Now, the isokhronon modification only applies to the constitutive law, and via the formalism developed, has the immediate implication of producing a background $X_a,\mathrm{D}X_a=0$, so it does not require further rederiving. Instead, let us look at self-interactions. Generally, the first order form of Proca theory just appends a Proca mass term to first order Maxwell theory. Equivalently, this can be seen as the first order form of massive Kalb-Ramond theory. Starting with the so-called parent or master path integral for Proca-Kalb-Ramond
\begin{equation}\label{eq:PKB_parent}
	Z_\text{PKB}=\int\mathcal{D}A\ \mathcal{D}B \exp\bigg[i\int-\frac{g^2}{2}B\wedge*B + A\wedge\mathrm{d}B + \frac{m^2}{2} A\wedge*A\bigg],
\end{equation}
either the vector potential $A$ or the auxiliary $B$ can be integrated out using basic path integral Gaussian formulas, resulting in either standard Proca theory or standard Kalb-Ramond theory
\begin{subequations}\label{eq:PKB_duality}
	\begin{align}
		Z_{\text{PKB}\setminus A}&=\int\mathcal{D}B\exp\bigg[i\int\frac{1}{2m^2}\mathrm{d}B\wedge*\mathrm{d}B - \frac{g^2}{2}B\wedge*B\bigg],\\
		Z_{\text{PKB}\setminus B}&=\int\mathcal{D}A\exp\bigg[i\int-\frac{1}{2g^2}\mathrm{d}A\wedge*\mathrm{d}A + \frac{m^2}{2}A\wedge*A\bigg].
	\end{align}
\end{subequations}
Alternatively, a field shift by a solution to the classical equations of motion decouples the fields, leading to the same conclusion. Such linear shift decoupling is \emph{not} possible for power-law terms higher than quadratic order, as the binomial coefficients will not match. In other words, it is not possible to eliminate interactions.

All of the constituents, that is the first order from of Proca and Kalb-Ramond and the theories themselves, \emph{must} be in agreement with each other: integrating out, as in the usual duality procedure~\cite{Hjelmeland:1997eg, Quevedo:1997jb}, manipulates the \emph{same} theory, and does not change anything in the basic physical premise, as it is ultimately just rewriting the mathematics in a different, closed form. More precise aspects of duality are a separate question, as it requires tracing how exactly the variables and the degrees of freedom or constants map to each other, but follows from the theory's definition. Indeed, Proca to Kalb-Ramond duality has long been a classical result.

A curious issue was brought up about self-interactions. Namely, it was found that the behaviour of transverse and longitudinal modes of the quartically self-interacting modifications of Proca and Kalb-Ramond differs in the massless limit~\cite{Hell:2022}. Specifically, there is a discontinuity in the massless limit of the quartically self-interacting theories, sourced from longitudinal modes for quartic Proca, while from transverse modes for quartic Kalb-Ramond. The Vainshtein scale is shared, but provided a non-zero mass, the propagating modes beneath the Vainshtein scale are different. Similar issues persist for other self-interacting $p$-form theories. Then, Ref.~\cite{Hell:2022} argues that this runs contrary to the duality of the \emph{basic} Proca and Kalb-Ramond theories themselves.

However, let it be emphasized that it is not clear why the \emph{self-interacting} theories should also be dual to each other. For example, the addition of Proca mass loses the self-duality of Maxwell theory, and instead suggests correspondence to Kalb-Ramond theory, leading to a qualitatively different situation, and this issue, as will be shown, also persists in the quartic case. Therefore, it is not clear what the limits of non-dual theories should imply about the duality of the basic theories. In fact, it would appear that quartic Proca and quartic Kalb-Ramond are \emph{not} dual to each other. Indeed, it was similarly noted in~\cite{Barbosa:2022zfm} that there does not appear to be any parent action which would produce both quartic Proca and quartic Kalb-Ramond, and duality only applies to the initial theories.

Let us add further that there are parent path integrals to produce quartic Proca and quartic Kalb-Ramond separately, but they do \emph{not} integrate to each other. Thus, quartic Proca and quartic Kalb-Ramond cannot be dual to each other. The path integral
\begin{equation}\label{eq:Proca4_parent}
	Z_{\text{Proca}^4}=\int\mathcal{D}A\ \mathcal{D}B\exp\bigg[i\int-\frac{g^2}{2}B\wedge*B+A\wedge\mathrm{d}B+\frac{m^2}{2}A\wedge*A + \lambda*(*(A\wedge*A))^2\bigg]
\end{equation}
integrates over $B$ to quartic Proca, so it must be a parent action for it, but there is no simple closed form solution to the quartic integral over $A$. Such integrals do not generally even make sense without renormalization. It can, however, be directly exemplified that~\eqref{eq:Proca4_parent} \emph{cannot} evaluate to (massive) quartic Kalb-Ramond
\begin{equation}\label{eq:quartic_KB}
	Z_{\text{KB}^4}=\int\mathcal{D}B\exp\bigg[i\int \frac{1}{2m^2}\mathrm{d}B\wedge*\mathrm{d}B -\frac{g^2}{2}B\wedge*B + \lambda*(*(B\wedge*B))^2\bigg].
\end{equation}
Separating terms in~\eqref{eq:Proca4_parent}, the path integral necessary for quartic duality does not generally hold,
\begin{equation}\label{eq:A2A4}
	\begin{aligned}
		Z_{A^2+A^4}[B]&=\int\mathcal{D}A\exp\bigg[i\int A\wedge\mathrm{d}B+\frac{m^2}{2}A\wedge*A + \lambda*(*(A\wedge*A))^2\bigg]\\
		&\neq\exp\bigg[i\int \frac{1}{2m^2}\mathrm{d}B\wedge*\mathrm{d}B + \lambda*(*(B\wedge*B))^2\bigg].
	\end{aligned}
\end{equation}
The second line consists of the additional terms required for~\eqref{eq:quartic_KB} to hold, while the first line is essentially the generating functional of correlation functions of the field strength $\mathrm{d}A$ of a theory with no dynamics. Evaluated at closed 2-forms $B=\mathrm{d}b$, explicitly
\begin{equation}
	Z_{A^2+A^4}[\mathrm{d}b]=\int\mathcal{D}A\exp\bigg[i\int\frac{m^2}{2}A\wedge*A + \lambda*(*(A\wedge*A))^2\bigg]
		\neq\exp\bigg[i\int \lambda*(*(\mathrm{d}b\wedge*\mathrm{d}b))^2\bigg].
\end{equation}
The left hand side loses dependence on $b$, while the right hand side does not, thus such an integration is not possible.

It is possible to define another auxiliary field $a\sim *(A\wedge*A)$, such that the action is at most quadratic in $A$, as in
\begin{equation}
	Z=\int\mathcal{D}A\ \mathcal{D}B\ \mathcal{D}a\exp\bigg[
	i\int-\frac{g^2}{2}B\wedge*B+A\wedge\mathrm{d}B+\frac{m^2}{2}A\wedge*A + \lambda a A\wedge*A + \lambda*a^2
	\bigg].
\end{equation}
Starting from this, formal integration over $A$ is easier, but integrating the resulting ``semi-dual'' theory over $a$ would still not be feasible. It does, however, imply a nontrivial interaction structure. More explicitly, quartic integrals have been suggested to evaluate through the full set of $n$-point correlation functions~\cite{Frasca:2015yva}. For scalar theory, also note the $(d-2)$-form ``notoph'' field theory~\cite{Ogievetsky:1966eiu}, which does not even admit non-derivative interactions, e.g. of $\phi^4$ type; see also~\cite{Mkrtchyan:2019opf}. Altogether, it appears that duality properties are not generally carried through interactions or modifications, and quartic self-interactions do not have a quartic correspondent in the dual theory. This is ultimately unsurprising, as the possible (self-)couplings can vary wildly for different basic fields.

\subsection{Maxwell-Chern-Simons}
\label{subsec:MCS}

Dimensional reduction aligns closely with the non-covariant Hamiltonian formalism utilized earlier. Doing so, the topology is assumed nontrivial, e.g. $M=C\times\Sigma$ with compact $C$ rather $\mathbb{R}\times\Sigma$ with temporal $\mathbb{R}$, and a lower-dimensional system is obtained when $C$ is shrunk to zero. In particular, e.g. $M=S^1\times\Sigma$ when compactifying by a circle, and the fields can be understood to situate on $\Sigma$ via the pullback by the embedding. As the one-dimensional $C$ admits a vector field $n$, this is realized through an analogy of the $3+1$ split discussed earlier. Note that the longitudinal and transversal components become separate fields. In the zero limit, this results in a field theory defined by the Lagrangian on the lower-dimensional manifold $\Sigma$. Lie derivative terms in the direction of $S^1$ should be handled carefully: in the simplest case, they can be simply deleted. Throughout this section, we will implicitly be working on a three-dimensional manifold.

Recently, dimensional reduction of first order theory was utilized to find the $S$-dual of Maxwell-Chern-Simons theory in three dimensions~\cite{Armoni:2022xhy},
\begin{equation}
	Z_\text{MCS}=\int\mathcal{D}A\exp\bigg[i\int-\frac{1}{2g^2}\mathrm{d}A\wedge*\mathrm{d}A + \frac{k}{4\pi}A\wedge \wedge\mathrm{d}A\bigg].
\end{equation}
In principle, this was done through a field shift of the parent path integral
\begin{equation}\label{eq:MCS_parent}
	Z_\text{MCS parent}=\int\mathcal{D}A\ \mathcal{D}b\exp\bigg[i\int-\frac{g^2}{2}b\wedge*b + b\wedge\mathrm{d}A+\frac{k}{4\pi}A\wedge\mathrm{d}A\bigg]
\end{equation}
by a solution to the equations of motion
\begin{equation}
	A\to-\frac{2\pi}{k}b + a,
\end{equation}
Here $a$ is the new deviation field, and $b$ is the ``magnetic'' 1-form. In principle, $b$ can be thought as to be postulated, or to originate from the $3+1$ split of the initial auxiliary 2-form $B$ in the 4-dimensional case. So, the dual partition function to Maxwell-Chern-Simons is
\begin{equation}\label{eq:MCS_dual}
	Z_\text{MCS dual}=\int\mathcal{D}a\ \mathcal{D}b\exp\bigg[i\int-\frac{g^2}{2}b\wedge*b-\frac{\pi}{k}b\wedge\mathrm{d}b + \frac{k}{4\pi}a\wedge\mathrm{d}a\bigg],
\end{equation}
and satisfies all requirements.

Most crucially, let us add that the same result can be derived from any basic parent path integral, through the direct integration procedure, provided the vacuum energy is handled carefully. The implication that dual theories have to coincide in the vacuum is rather subtle and fundamental. Specifically,~\eqref{eq:MCS_parent} \emph{already} admits the Gaussian integration formula
\begin{equation}
	\int\mathcal{D}A\exp\bigg[i\int A\wedge \mathrm{d}b+\frac{k}{4\pi}A\wedge\mathrm{d}A\bigg]\sim\exp\bigg[-\frac{i}{2}\int\mathrm{d}b\wedge\bigg(\frac{k}{2\pi}\mathrm{d}\bigg)^{-1}\mathrm{d}b \bigg]\frac{1}{\sqrt{\det((k/2\pi)\mathrm{d})}}.
\end{equation}
Although the inverse of the exterior derivative is rather formal, it can, to some extent, be understood in the sense of the linear homotopy operator~\cite{Edelen:1985}. Loosely, in a star-shaped region $S$ around the point $P_0(x_0^i)$, with the adapted coordinate system
\begin{equation}
	\tilde{\alpha}(\lambda)=\alpha_{i_1\ldots i_p}(x_0^j + \lambda(x^j - x_0^j))\mathrm{d}x^{i_1}\wedge\ldots\wedge\mathrm{d}x^{i_p}
\end{equation}
and radius vector field
\begin{equation}
	\mathcal{X}(x^i)=(x^i-x_0^i)\partial_i,
\end{equation}
the linear homotopy operator $H:\Omega^p(S)\to\Omega^{p-1}(S)$ is defined as the Riemann-Graves integral
\begin{equation}
	H\alpha=\int_0^1\mathcal{X}\lrcorner\tilde{\alpha}(\lambda)\lambda^{p-1}\mathrm{d}\lambda.
\end{equation}
Of particular importance is $\mathrm{d}H+H\mathrm{d}=\mathds{1}$ for $p\geqslant1$, but currently, $\mathrm{d}^{-1}$ only serves to eliminate $\mathrm{d}^{-1}\mathrm{d}=\mathds{1}$, so it is possible that explicit inversion can be bypassed. Most importantly, note that the functional determinant term is the propagator of a decoupled free Chern-Simons theory
\begin{equation}
	\frac{1}{\sqrt{\det((k/2\pi)\mathrm{d})}}\sim\int\mathcal{D}b\exp\bigg[\frac{k}{4\pi}b\wedge\mathrm{d}b\bigg].
\end{equation}
The evaluation results in~\eqref{eq:MCS_dual}, as required.

In principle, integrating the path integral changes nothing, except for how the physical content is represented, so the dual theory~\eqref{eq:MCS_dual} is unsurprising. However, as we have found, \emph{all} of the physical content is critical, even when relating only to the vacuum. As noted~\cite{Armoni:2022xhy}, only this full dual theory satisfies all of the requirements of a dual theory, e.g. global $\mathbb{Z}_k$ symmetry, pure Chern-Simons in the infrared, etc. Note that the Chern-Simons case is particular, as the integration yields a nontrivial functional determinant. In the pure Proca case considered in section~\ref{subsec:Proca_Kalb_Ramond}, the field shift decouples an entirely non-dynamical contribution, e.g. $\frac{m^2}{2}a\wedge*a$, which would provide only an indeterminate path integral of a non-dynamical theory, not relevant for most physical inquiries, but still implicitly retained. Similarly to other topological modes or decoupled fields, it often is not of physical interest. But all of this does suggest understanding duality as just different facets of a single (quantum) system, rather than necessarily a correspondence between different theories. The rewriting is particularly obvious in the path integral formalism.

Let us add that the Proca-Chern-Simons case does not admit a closed form solution. The parent action appends the 3-dimensional Proca mass
\begin{equation}
	Z_\text{PCS parent}=\int\mathcal{D}A\ \mathcal{D}b\exp\bigg[i\int-\frac{g^2}{2}b\wedge*b + b\wedge\mathrm{d}A+\frac{m^2}{2}A\wedge*A+\frac{k}{4\pi}A\wedge\mathrm{d}A\bigg],
\end{equation}
and the ``magnetic'' dual would require integrating over or decoupling $A$. Solving the equations of motion
\begin{subequations}
	\begin{align}
		\mathrm{d}b&=-\bigg(\frac{k}{2\pi}\mathrm{d}+m^2*\bigg)A,\label{eq:EoM_PCS_A}\\
		g^2*b&=\mathrm{d}A
	\end{align}
\end{subequations}
inevitably requires inverting a differential operator. It is possible to write a formal infinite series
\begin{equation}
	A=\bigg(\mathds{1}+\frac{k}{4\pi m^2}*\mathrm{d}+\frac{k}{4\pi m^2}*\mathrm{d}\frac{k}{4\pi m^2}*\mathrm{d}+\ldots\bigg)\frac{1}{2m^2}*\mathrm{d}b,
\end{equation}
although convergence is a severe question, and neither would an attempt at some Borel-like transformation clarify this. Nonetheless, see~\cite{Kycia:2024} for a proof that a similar power series solution to $\mathrm{D}\alpha=0$ is indeed convergent\footnote{This also allows writing a formal solution to the isokhronon background $X_a,\mathrm{D}X_a=0$, although actual evaluation remains difficult.}; see~\cite{Kycia:2022} with respect to the codifferential and Dirac-Kähler fields.

Nevertheless, it is possible to write down an implicit dual theory to Proca-Chern-Simons. A field shift
\begin{equation}
	A\to\tilde{b} + a
\end{equation}
to a deviation around a solution $\tilde{b}=\tilde{b}(b)$ such that
\begin{equation}
	\mathrm{d}b=-\bigg(\frac{k}{2\pi}\mathrm{d}+m^2*\bigg)\tilde{b},
\end{equation}
decouples
\begin{equation}
	Z_\text{PCS dual}=\int\mathcal{D}a\ \mathcal{D}b\exp\bigg[i\int-\frac{g^2}{2}b\wedge*b
	- \frac{m^2}{2}\tilde{b}\wedge*\tilde{b}
	- \frac{k}{4\pi}\tilde{b}\wedge\mathrm{d}\tilde{b}
	+ \frac{m^2}{2}a\wedge*a
	+ \frac{k}{4\pi}a\wedge\mathrm{d}a\bigg].
\end{equation}
By the preceding discussion, this must define the ``magnetic'' dual theory of 3-dimensional Proca-Chern-Simons, although it inevitably remains in an implicit form, and as the decoupled Chern-Simons term retains a derivative component, further integration is not even allowed.

\subsection{Duality rotations}
\label{subsec:duality_rotations}
Let us conclude with duality transformations. The question becomes of transforming (rotating) one component of dual theories to another. Duality rotations between electric and magnetic variables are well studied, and it is known that these do not generalize to non-Abelian gauge theory in the basic variables~\cite{Deser:1976iy}. Explicit duality invariance of nonlinear electromagnetism in the action requires modification, like auxiliary variables or losing explicit Lorentz invariance, see also~\cite{Avetisyan:2021heg} and the references therein. Let us show that the non-Abelian issue is immediate when considering the differential geometry of the situation.

The vacuum equations of motion
\begin{equation}
	\mathrm{D}*F=0
\end{equation}
are preserved by a formal symmetry rotation
\begin{equation}\label{eq:similarity_rotation}
	F\to\cos\alpha\ F + \sin\alpha\ {*F},
\end{equation}
where the angle $\alpha$ is a global constant. Note that the Bianchi identity $\mathrm{D}F=0$ is satisfied by construction for any field strength 2-form and does not carry additional information. The rotation~\eqref{eq:similarity_rotation} is not a Noether symmetry of the action
\begin{equation}\label{eq:basic_YM_action}
	S_\text{YM}=\int\Tr_G\bigg(\frac{g^2}{2}F\wedge*F\bigg).
\end{equation}
which instead transforms to
\begin{equation}\label{eq:duality_transformed_action}
	S'_\text{YM}=\int\frac{g^2}{2}\Tr(\cos(2\alpha)F\wedge*F-2\sin\alpha\cos\alpha\ F\wedge F),
\end{equation}

The basic geometric and physical issue, as will be shown, is that a (local or global) scalar multiple of curvature (i.e.\ field strength) is not itself necessarily the curvature of some other connection (i.e.\ gauge field). The implication is that the space of curvature 2-forms is not continuously spanned by scalar multiples.  That is, curvatures do not form a homogeneous space. Thus, there is no generally consistent way to produce the duality rotation~\eqref{eq:similarity_rotation} via a transformation of the connection, recovering the results of~\cite{Deser:1976iy}.

Contrariwise, if a global rotation could be induced by a transformation of the basic connection variable, it would require that
\begin{equation}\label{eq:rescaled_curvature}
	F[B]=a F[A]
\end{equation}
be solvable for some new connection 1-form $B$ and an arbitrary (global) rotation coefficient $a$. However, solving the nonlinear differential equation
\begin{equation}\label{eq:rescaled_curvature_difeq}
	\mathrm{d}(B - a A) + B\wedge B - a A\wedge A=0
\end{equation}
is generally not possible. More specifically, the solvability depends on the dimension of spacetime and the gauge group. In particular, Abelian gauge groups have $A\wedge A=B\wedge B=0$, and the question is of finding a potential for a closed form. But generally, for arbitrary non-Abelian theory, this is not possible.

To see the issue, note the Bianchi identity
\begin{equation}
	\mathrm{D}_B F[B]=a\mathrm{D}_B F[A]=0\Rightarrow \mathrm{d}F[A]=-[B, F[A]]
\end{equation}
transforms the differential equation into a system of linear algebraic equations\footnote{Due a related method by R. Bryant. See the general study of exterior differential systems~\cite{Bryant_et_al:1991}, as ultimately this is a question of the solvability of certain differential equations on manifolds.}. Also note that $\mathrm{d}F[A]=-[A, F[A]]$ implies the homogeneous form,
\begin{equation}
	[A-B,F[A]]=0\sim f_{IJ}{}^K F[A]^I\wedge(A^J - B^J)=0.
\end{equation}
Defining the shift as the new variable $C^I=A^I-B^I$ yields in matrix form
\begin{equation}
	\begin{pmatrix}
		f_{I1}{}^1 F[A]^I 	&f_{I2}{}^1 F[A]^I 	&	\cdots\\
		f_{I1}{}^2 F[A]^I 	&f_{I2}{}^2 F[A]^I 	&	\cdots\\
		\vdots 						&\vdots 						&
	\end{pmatrix}
	\wedge
	\begin{pmatrix}
		C^1\\C^2\\\vdots
	\end{pmatrix}
	=0,
\end{equation}
which can, in turn, be further expanded in components $C^I = C^I{}_\mu\mathrm{d}x^\mu$ for a linear homogeneous system for functions over spacetime. A particular solution is $C=0$, that is $B=A$, which is inconsistent with~\eqref{eq:rescaled_curvature}. The existence of nontrivial solutions would require the system matrix not to be invertible, i.e.\ the rank of $f_{IJ}{}^K F[A]^I$ to be less than maximal, which is difficult to achieve as $F[A]^I$ are independent from each other. As a solution to a linear system, the components of $B$ are necessarily a linear composition of $A$ and $F[A]$. The general case of a nontrivial kernel for $[C,F[A]]=0$ requires gauge group and spacetime-specific handling. A local coefficient $a(x)$ would not change the issue.

Alternatively, note the Hodge decomposition on a closed manifold
\begin{equation}
	B\wedge B - a A\wedge A = \mathrm{d}(\alpha[B\wedge B] - a\alpha[A\wedge A])+\delta(\beta[B\wedge B] - a\beta[A\wedge A]) + \gamma[B\wedge B] - a\gamma[A\wedge A],
\end{equation}
where the terms are correspondingly exact, co-exact and harmonic, i.e.\ $(\mathrm{d}\delta + \delta\mathrm{d})\gamma[\cdot]=0$. After parameter and variable redefining, this yields the equivalent system
\begin{subequations}
	\begin{align}
		\mathrm{d}(\alpha[B\wedge B] - \alpha[A\wedge A])&=\mathrm{d}(B - b A),\label{eq:rotation_Hodge_a}\\
		\delta(\beta[B\wedge B] - \beta[A\wedge A])&=0,\label{eq:rotation_Hodge_b}\\
		\gamma[B\wedge B] - \gamma[A\wedge A]&=0.\label{eq:rotation_Hodge_c}
	\end{align}
\end{subequations}
The operators of the exact $\alpha[\cdot]$, co-exact $\beta[\cdot]$ and harmonic part $\gamma[\cdot]$ do not have simple properties. They are, however, linear operators, because of the direct sum nature of the Hodge decomposition. Equations~\eqref{eq:rotation_Hodge_b} and~\eqref{eq:rotation_Hodge_c} have $A=B$ as a particular solution, which is generally inconsistent with~\eqref{eq:rotation_Hodge_a}. In another vein, as they are linear operators, any two solutions $B\wedge B$ must differ by an element of the kernel. In particular, this means that
\begin{equation}
	B\wedge B=A\wedge A + C_1,\ \beta[C_1]=0\text{ or }\gamma[C_1]=0.
\end{equation}
But returning to~\eqref{eq:rotation_Hodge_a} this implies that
\begin{equation}
	B=b A + \alpha[C_1] + \mathrm{d}C_2,
\end{equation}
which is, generally, again inconsistent with~\eqref{eq:rescaled_curvature}. In particular, note the Bianchi identity
\begin{equation}
	\mathrm{D}_A F[b A + \alpha[C_1] + \mathrm{d}C_2]=0
\end{equation}
is not trivially satisfied, and would instead impose a new condition on the otherwise mostly arbitrary 2-form $C_1$ and 0-form $C_2$.

Furthermore, the differential geometry of the dynamical origin of the issue is straightforward. For Abelian groups, the curvature $F=\mathrm{d}A$ is exact (thus closed) and its Hodge dual $*F$ is co-exact (thus co-closed). So desired, $*F$ can be seen as generated by a Hodge dual potential,
\begin{equation}
	*F=\delta*A.
\end{equation}
Off-shell, this is also the limit of duality: it is not possible, generally, to say whether $*F$ has some well-defined vector potential, and trying to rotate $A$ into $*A$ does not have a direct meaning other than the Hodge pairing itself.

However, the equations of motion $\mathrm{d}*F=0$ give the nontrivial implication that $*F$ is \emph{also} closed, thus by the Poincaré lemma, it is also exact, and there is a well-defined potential such that $*F=\mathrm{d}B$. In the context of path integrals, this is the result of integrating the coupled term $A\wedge\mathrm{d}B$ to a Dirac delta $\delta(\mathrm{d}B)$. Because the field strength is linear in the Abelian connection, the formal similarity rotation~\eqref{eq:similarity_rotation} becomes a proper rotation of the gauge field, i.e.\ the connection 1-form $A$, to produce the dual field strength $*F$,
\begin{equation}
	A\to\cos\alpha\ A + \sin\alpha\ B,
\end{equation}
and the system of equations of motion is preserved. The parameter $\alpha$ is to be global for a rotation in $F$, see also~\cite{Deser:2010it}.

There are insurmountable difficulties when moving to non-Abelian theory. The equations are nonlinear, and the system loses the cohomological utility. There is no Poincaré lemma for the exterior covariant derivative, in particular $\mathrm{D}_A^2\sim F[A]\wedge\cdot\neq0$, and there is no de Rham cohomology. So, $F=\mathrm{d}A+A\wedge A$ is \emph{covariantly} closed, $\mathrm{D}_A F=0$ the Bianchi identity, and $*F$ covariantly co-closed, $\delta_A*F=0$. Neither have a sense of exactness, $F\neq\mathrm{D}_A A$ and $*F\neq\delta_A A$, and covariant constancy does not necessarily imply proportionality to curvature. So the prototype Bianchi identity, resp. the vacuum equation of motion, $\mathrm{D}_A*F=0$ does not imply that $*F$ is necessarily a curvature of some other magnetic connection $B$, nor would it make sense, as $\mathrm{D}_A*F=0$ would be the Bianchi identity with respect to the initial connection $A$, rather than some other $B$.

Altogether, there are several immediate geometric obstructions against implementing duality rotations in the basic variables for non-Abelian theories. The cohomological apparatus does \emph{not} carry over to the exterior covariant derivative, and the nonlinear nature of curvature forbids most any transformation, outside degenerate cases. The duality rotations only persist as formal \emph{similarity} rotations in the general case. Recently there was a disproof of $U(1)$-type duality rotations in general relativity~\cite{Monteiro:2023dev}. Let us note that the formalism here also extends to gravity when thought of as a constrained BF theory, with the Lorentz connection in the role of the vector potential. The simplicity constraints are more subtle, but do not change the premise.

A weaker sense of duality persists, but does not appear to be practically useful. The transformation~\eqref{eq:similarity_rotation} is a mapping from the space of curvatures to general 2-forms. The transformed action~\eqref{eq:duality_transformed_action} is \emph{physically} equivalent to the initial~\eqref{eq:basic_YM_action}, up to a topological term and a global constant. Neither contribution is generally physically distinguishable. So, it is pertinent to introduce a physical equivalence relation $\sim$, and claim either form to be simply representatives of the same equivalence class. More nuanced interpretations can define equivalence up to some observables, while a particularly simple definition would collect all actions which provide the same system of Lagrangian equations to the same class. The Noether procedure retains its form, while various objects also become denoters for equivalence classes.

\section{Conclusion}
\label{sec:conclusion}
The focus of interest was in studying the consistency of the isokhronon gauge theory~\cite{Gallagher:2022kvv}. Although the analysis was for electromagnetism, Yang-Mills theory should not be significantly different if the trace is handled consistently. As it currently appears, the gauge theory manages to satisfy a substantial analogy with the khronon gravity theory it was inspired from, and it manages to exhibit a variety of curious properties. It does not manage to be physically fully satisfactory, alas, at least not in its current form. Nevertheless, now there is a Hamiltonian study available to complement the Lagrangian understanding of the auxiliary isokhronon vector $\phi^a$ gauge theory: as long as the torsionful background is compatible, the theory behaves as electromagnetism with a constant vacuum excitation background would, both on a classical and quantum level. In principle, the methods and arguments carry over to other constitutive laws with differential equations.

The canonical structure is surprisingly similar to its analogous gravitational theory~\cite{nikjoo2023hamiltonian}. The count of the degree of freedom diverges, however, as here the Dirac-Bergmann algorithm does not appear to suggest new local degrees of freedom, while the khronon Hamiltonian establishes cosmic dust degrees~\cite{Zlosnik:2018}. This isokhronon vector gauge theory is a curious example of how derivatives in the constraints behave in the Dirac-Bergmann algorithm; the consistency of the algorithm for differential constraints was recently questioned in~\cite{dambrosio:2023}. The constraints exhibit a peculiar form of intermediary spontaneous Lorentz symmetry breaking. This theory is an example of intrinsic backgrounds, thus a parallel to the background freedom in~\cite{kaplan:2023a, kaplan:2023b}. As discussed, this relates to the fundamental question of background independence of \emph{all} physical theories.

Applications to duality, in turn, are immediate, and the differential geometric apparatus proves useful. The path integral agrees with symplectic arguments. Indeed, as found, the parent path integral is to be invariant in deriving dual theories. This shows that quartic Proca is not dual to quartic Kalb-Ramond, so limit claims~\cite{Hell:2022} are not necessarily consistent. This invariance of the parent path integral is also a crucial implication from Maxwell-Chern-Simons theory~\cite{Armoni:2022xhy}, and applied to Proca-Chern-Simons, really implies a consistent handling of the vacuum, as is found. The differential geometry of duality rotations is also immediate, with the Maxwell (and any Abelian) case leaning on the cohomology of the exterior derivative, resp. the Poincaré lemma, while non-Abelian Yang-Mills theory does not admit such results. The scalar multiple of a curvature is not necessarily the curvature of another connection, and there is no Poincaré lemma for the covariant derivative, giving a global description to~\cite{Deser:1976iy}.

The standard Hamiltonian analysis and the Dirac-Bergmann algorithm is straightforward and well studied, in its requirements and implications. Isokhronon electromagnetism provides an awkward example of the subtleties involved, but it shouldn't prove too difficult to invent similar, better and simpler models if the pregeometry and analogy requirements were to be relaxed. Nevertheless, the results here are relatively clear: the theory reduces to electromagnetism with a vacuum excitation. The degree of freedom count implies two propagating polarizations and a global excitation degree, with the constraint algebra consisting of the momentum definitions, Gauss law and an additional secondary constraint. All constraints are differential equations, which introduces the background excitation ambiguity, mediated by the auxiliary vector's momentum. This behaviour is similar to the gravity theory~\cite{Zlosnik:2018}, except for not providing the background as an additional degree of freedom. This allows to conclude about both theories: the anti-self-dual component of the self-dual khronon gravity theory is vital to the model. The isokhronon gauge theory lacks such structure, especially so in its electromagnetic variant.

In addition, the secondary constraint exhibits what appears to be an odd variant of intermediary spontaneous Lorentz symmetry breaking. Solving for the constraint requires choosing an explicit, arbitrary spacetime direction. For comparison, the spontaneous symmetry breaking in khronon gravity manifests as the titular khronon vector obtains a non-zero expectation value. The difference in constraint structure and behaviour is not entirely unsurprising, however, as the independent excitation (or, field strength) of first order gauge theory fare lacks the tetrad or coframe interpretation of the similar terms in gravity. The mathematical structure is simply different, see~\cite{Weatherall:2016} for more discussion. In a very basic way, first order electromagnetism introduces an auxiliary equation to integrate out the independent field excitation (with possibly additional effects, as here), while for first order gravity the metric (or tetrad, coframe etc.) defines the basic gravitational equations. Note that, classically, this is still an intermediary process, if the initial data is permitted to transform covariantly. Although there can be a preferred frame keeping the background $X_a$ invariant, this is not a strict implication.

The noncovariant analysis appears rather complicated. In contrast, the covariant phase space is simpler, but significantly more implicit. The covariant phase space is constituted by the solutions of the equations of motion, modulo zero modes' gauge transformations, but does not provide a strictly explicit way to describe the phase space and constraints, bar solving the equations of motion explicitly. There is an inherent necessity for other (Lagrangian) machinery to have a comprehensive description of the geometry, the solutions, or the degree of freedom count. What is important, however, is the shift in the geometric understanding, and relating the variation to the symplectic structure. The conserved currents follow, while the symplectic structure can be seen as the basis for quantization. In addition to various minor addenda to the procedures, the covariant phase space allows to characterize the quantum isokhronon gauge theory, while also giving a hint at what the (loop, or other) quantization of the khronon gravity theory might provide. Ultimately, however, the covariant phase space remains a difference in description.

The background torsion requirement of the isokhronon theory is either concerning (to the point of dismissing), or arguable (up to continuing some investigation), particularly in conjunction with khronon gravity. For now, it would appear if the background is compatible with the auxiliary vector, then the theory is consistent, and reduces to electromagnetism with an integration constant background. Nevertheless, it will be important to verify that this is actually true in the coupled gravity-gauge theory --- that is, whether its accompanying gravitational theory can provide a consistent torsion background throughout.

As mentioned, there is a more fundamental physical question of background independence. Here this is expressed by an embedding ambiguity in certain derivative forms of the gauge theory constitutive law. Kaplan et al.~\cite{kaplan:2023a, kaplan:2023b} argue that the Hilbert space of gauge states can logically be made larger on gauge invariance grounds alone, with states including an additional electromagnetic background. The gauge theory considered here, in addition to its gravitational analogue~\cite{Zlosnik:2018}, provide similar effects both on a classical and, as argued, on a quantum level. The most explicit background addition would add an arbitrary current $A\wedge J$ (or excitation $A\wedge\mathrm{d}H$), forced to transform such that the action remains gauge invariant, cf. Noether's second theorem and the conservation equation. The question is in the logical completion.

All such modifications could be dismissed if taking background independence as a fundamental principle, required for all physical theories. In addition, there are other questions, such as background quantum transitions, background dynamical principles or evolution equations (should it be completely independent from matter, or suffer some backreaction?), issues of arbitrariness in background choice, to not mention the Standard Model precision tests which have not suggested any such phenomena. The basic issue is that Nature does not appear to provide any notably strong anomalous backgrounds, especially so in a laboratory microscopic (as opposed to cosmological) scale. This suggests that even if such backgrounds were allowed, they should somehow be restricted in their strength, fluctuation and distribution. If there's some tempora-spatial scope of applicability of quantum field theory (rather than extending over the entire Universe in one gigantic quantum system), then it is rather awkward to have the background be compatible around different measurements.

In any case, background independence of a gauge theory here holds a rather concrete meaning in relation to background currents; for reference in gravitation, it is the independence of the defining equations from preferred backgrounds or coordinate systems. Nevertheless, the philosophical quandary remains. An optimist could hypothesize from this that perhaps the usual gauge theory language should be adapted in a way which does not even make such ambiguity possible. A purely canonical description of the phase space geometry of the polarizations would certainly hide any such questions. Also, dismissing backgrounds does not necessarily dismiss gauge theories which include a zero background possibility among nontrivial options --- then there is just a canonical zero background.

There was a fair amount of comment of pregeometric theory, and degenerate metrics in particular. Although the full spectrum of theory modification in this vein is impossible to constrain, one concrete aspiration was provided: to incorporate a consistent topological phase and transition. Let us also provide a concrete proposal how to study some effects, if any, of a class of metric degeneracies. It is apparent that such effects are of continuity and topological (contention) nature. Then, some insight can be derived from topologically nontrivial manifolds. In particular, if a topological hole can be contracted to a point in the sense of the outside metric, then the limits of relevant quantities describe how a degenerate zone would affect the metric region. Although this is a relatively natural continuation into singularities, such questions, if at all, only arise in extreme situations, like in a black hole singularity, or in curious theories.

Let us conclude in a forward-looking manner. Although the isokhronon gauge theory runs afoul of peculiar issues, it maintains quite strong similarity to the gravity theory it was meant to accompany, while simultaneously remaining rather distinct. There is, however, no immediately obvious or unique path to unification from these gauge theory grounds. The phenomenology is, with this paper, essentially clear; a question would be in characterizing covariantly constant forms (of the type $\mathrm{D}\alpha=0$), but a hypothesis would be it's simply extending values from a point in a connection-consistent manner; see also~\cite{savvidy2024large}. Indeed, this interpretation is also consistent with results derived from the linear homotopy operator~\cite{Kycia:2022,Kycia:2024}. Overall unification was discussed at more length earlier in~\cite{Gallagher:2022kvv}. It is not, however the only option forward, and other modifications might promise better perspectives.

Electromagnetism is unique in the sense that it is defined by an Abelian group. This makes it distinct from other relativistic field theory-wise relevant symmetry groups, and dismisses a deal of action candidates that hold better resemblance to gravity. So, it might be better to attempt graviweak unification~\cite{Nesti:2007ka} in first order theory, with inspiration from Plebanski gravity~\cite{Smolin:2007rx}. For non-Abelian gauge groups, it is possible to at least write down
\begin{equation}
	S=\int\Tr\bigg(\frac{1}{2} u\wedge u\wedge *(u\wedge u) + u\wedge u\wedge F\bigg),
\end{equation}
for some adjoint-valued $u\sim\mathrm{D}\phi$, in some similarity to~\cite{husain2023general} (cf.~\cite{Zlosnik:2018}), and with much greater analogy to gravity. It would be curious to verify whether modifications of such gauge theories possess a similar constraint structure in the Hamiltonian analysis, and whether the choice of a preferred direction could be operated for symmetry groups other than Lorentz symmetry, or whether the Hodge (or Lorentz, viz. the Levi-Civita symbol) dualization is crucial in the process. A different approach could proceed from studying constrained BF-theory. With a cosmological constant addition, it is indistinguishable from first order gauge theory, while gravity is unique in the simplicity constraints, so it would be curious to see how far this analogy can be pushed. The current work can provide some inspiration in this direction.

\begin{acknowledgments}
The author would like to thank T. S. Koivisto, T. Zlosnik and P. Jirou\v{s}ek for helpful discussions. This work was supported by the Estonian Research Council grant PRG356 ``Gauge Gravity'', the Estonian Research Council CoE program with the grant TK202 ``Fundamental Universe'' and by the EU through the European Regional Development Fund CoE program TK133 ``The Dark Side of the Universe''.
\end{acknowledgments}

\bibliography{references}

\end{document}